\documentclass[12pt,a4paper]{article}
\usepackage[utf8]{inputenc}
\usepackage[english]{babel}
\usepackage{amsmath}
\usepackage{amsfonts}
\usepackage{isomath}
\usepackage{amsmath}
\usepackage{amssymb}
\usepackage{amscd}
\usepackage{amsfonts}
\usepackage{fancyhdr}%per intestazione e piè di pagina
\usepackage{graphics}
\usepackage{verbatim}
\usepackage{tabularx}
\usepackage{subfigure}
\usepackage{xspace}
\usepackage{euscript}
\usepackage{alltt}
\usepackage{boxedminipage}
\usepackage{float}
\usepackage{color}
\usepackage{varioref}
\usepackage{hyperref}
\usepackage{tabularx}
\usepackage[all]{xy}
\usepackage{t1enc}
\usepackage{times,exscale}
\usepackage{graphicx,calc}
\usepackage{mdwlist}
\usepackage{framed}
\usepackage{enumitem}
\usepackage{stmaryrd}
\usepackage{mathrsfs}
\usepackage{mathtools}
\usepackage{multirow}

\usepackage[margin=20mm]{geometry}
\numberwithin{equation}{section}
\usepackage{booktabs}
\usepackage{multirow}
\usepackage{graphicx}
\usepackage[table,xcdraw]{xcolor}
\everymath{\displaystyle}
\date{$\quad$}
\usepackage[T1]{fontenc}
\usepackage{authblk}
\usepackage{cite}

\definecolor{red}{rgb}{1.0, 0.0, 0.0}
\definecolor{blu}{rgb}{0.0, 0.0, 255.0}
\usepackage[font=small,labelfont=bf]{caption} 

{\left\lbrace\begin{array}{@{}l@{}}}%
{\end{array}\right.}

\usepackage{comment}
\usepackage{stmaryrd}
\usepackage{cite}
\usepackage[table,xcdraw]{xcolor}
\usepackage{lineno}
\usepackage[bbgreekl]{mathbbol}
\usepackage{lineno}

\graphicspath{{IMG//}}

\title{\titlescript}
\newcommand{\titlescript}{On growth and morphogenesis in mechanobiology}
\author[1,2]{A.R. Carotenuto}
\author[1,2]{S. Palumbo}
\author[1,2]{A. Cutolo}
\author[1,2,3]{M. Fraldi}

\affil[1]{\small{Department of Structures for Engineering and Architecture, University of Napoli "Federico II" - Italy}}
\affil[2]{\small{Laboratory of Integrated Mechanics and Imaging for Testing and Simulation (LIMITS), University of Naples “Federico II”, Napoli, Italy;}}
\affil[3]{\small{LPENS - Départment de Physique, Ecole Normale Supérieure, Paris, France;}}

\graphicspath{{IMG//}}

\begin{document}
%\linenumbers
%
\pagestyle{fancyplain}
\lhead{\fancyplain{\scriptsize \titlescript }{\scriptsize \titlescript}}
\rhead{\fancyplain{\scriptsize 1}{\scriptsize ...}}
\maketitle

\begin{abstract}
Morphoelasticity represents a foundational theory for tracing back growth, remodelling, and morphogenesis, yet crucial challenges persist. A unified growth law —independent of \textit{a priori} assumptions about constitutive relations or specified structures of the growth tensor— remains in fact elusive, as does an intimate connection between local anisotropic growth, morphogenesis dynamics, diffusion phenomena and mechanics. When anisotropy of growth is not prescribed arbitrarily, current frameworks indeed mainly attribute shape emergence to growth-induced configurational switches at macroscopic scales, somehow neglecting the existence of local drivers of spontaneous patterning or distorsions also in stress-free conditions.

To overcome these limitations, in this work we propose a theoretical approach that reformulates anisotropic growth, remodeling, and morphogenesis starting from first principles, grounded in mass balance. In particular, by positing mass conservation as the sole dynamic constraint for the growth problem, we generalize the mass balance equation to derive evolution laws for mass distribution and geometric reconfiguration. This reveals how mass transport and associated microstructural evolution of the tissue fabric synergistically guide biological, geometric, and constitutive changes of living matter, with the isochoric component of growth orchestrating local anisotropy and, at the macroscale, even spontaneous shape development.

The proposed fully coupled modelling approach integrates chemo-mechanical and biological interactions inherent to living systems, offering a pathway to investigate complex scenarios in growth and morphogenesis, redefining the latter in a Continuum Mechanics framework. It is felt that this strategy could help to take a step forward a unified perspective for biomechanical and mechanobiological problems, bridging scales from local drivers to emergent tissue forms and shapes.
\end{abstract}
\section{Introduction}
\label{intro}
%Growth
Within the vivid scenario of mechanobiology, the variety of evolution phenomena undergone by living materials has been systematized into the three prominent pillars of growth, remodelling and morphogenesis. Although these processes often occur synergistically without the possibility of establishing sharp boundaries among the different biophysical aspects involved, some essential hallmarks are taken as reference for their phenomenological modelling in biomechanics. As well-known, macroscopic growth is basically characterized by a change of mass within the system \cite{goriely2017mathematics, lubarda2002mechanics, taber1995biomechanics, jones2012modeling}. Adaptation of the biological structure during growth, also in response to specific load conditions or external constraints, kindles remodelling. This provides changes in material properties or density (material remodelling), modified by the residual stresses and elastic strains imprisoned in the material \cite{Carotenuto_2019, Ambrosi_2006}, but can also occur as adaptation of microstructure (structural remodelling), which instead involves a macroscopic re-orientation of the material architecture with a consequent impact on overall strength and stiffness \cite{ cowin1992evolutionary, tilahun2023biochemomechanical, Hariton_2006, fraldi2022toughening, Goda_2016}.
%What is intended with morphogenesis: instabilities, diffusion, morphoelasticity etc.
On the other hand, morphogenesis is essentially recognized as a change of shape \cite{taber2020continuum, goriely2017mathematics}, which is traditionally associated to developmental growth rather than physiological processes. For instance, during embryogenesis, morphogenesis constitutes a relevant feature by influencing differentiation and spatial re-distribution of the living constituents by patterning the form under prescribed stimuli and constraints \cite{Ben_Amar_2013, dervaux2008morphogenesis, jones2012modeling}. However, in literature, various interpretations of morphogenesis have been provided with reference to specific biological processes. 
%%%
Within the most general framework, analogously to many other coupled multi-field mechanical theories \cite{Lubarda_2004}, the growth problem in biomechanics is formulated through the coupling of mechanical equilibrium equations involving e.g. the (nominal) stress $\mathbf{P}$ with balance equation of the mass $m$, which involves possible feedback terms with environmental stress and chemicals. Herein, the growth is considered as an inelastic deformation by means of the canonical multiplicative decomposition of the total deformation gradient $\mathbf{F}$ into a growth part $\mathbf{F}_g$ and an elastic part $\mathbf{F}_e$. This latter one is the sole responsible of mechanical stresses, which emerge as a consequence of the accumulation of a certain strain energy density $\psi$ according well-established constitutive frameworks and can be induced either by external loads or internal adaptation due to  general growth incompatibility. Such a kinematic and constitutive assumptions represent, together with conservation equations, the well-established theory of volumetric morphoelasticity \cite{goriely2017mathematics}, according to which the growth problem can be written with respect to a selected reference state $\mathcal{B}_0\subset \mathbb{E}^3$ as: 

\begin{equation}\label{0prob}
\begin{cases}
    &\nabla_0\cdot\mathbf{P}=-\textbf{B}\\
    &\dot{m}=f(m,\mathbf{P}-\mathbf{P}^*, c_i; \mathbf{X},t)\\
    &\mathbf{F}=\mathbf{F}_e\mathbf{F}_g\\
    &\mathbf{P}= J_g\left[\frac{\partial\psi}{\partial\mathbf{F}_e}\right]\mathbf{F}_g^{-T}, \quad J_g=\det\mathbf{F}_g
\end{cases}
\end{equation}

This system represents a closed problem if one \textit{a priori} assumes both the structure of the growth tensor $\mathbf{F}_g$ as well as its relation with the mass growth, in the most of cases expressed in terms of purely volumetric growth (by assuming that bulk density changes are exclusively connected to elastic strains). Since this freedom, the spatial growth problem might result paradoxically not unique in conceiving the way in which mass variation produces geometric variation within the body.
In fact, since the total geometric changes depend also on the form of the local inelastic growth tensor, in this type of approach there is the need to introduce constitutive assumptions for $\mathbf{F}_g$. For instance, the $\mathbf{F}_g$ structure can be specified considering diagonal tensors in which the total volumetric growth is partitioned along the three principal directions either in isotropic way or anisotropically, depending on anisotropy coefficients that can be either assigned or, in fully coupled models, oriented by the stresses \cite{katsamba2020biomechanical, carotenuto2021lyapunov, lamm2022macroscopic, kuhl2014growing, ben2013anisotropic}. In these models, the role of the growth Jacobian is well defined and, if densification/rarefaction is excluded in the natural state, there is formal correspondence between mass and volume change. However, to the best of Authors' knowledge, less attention is focused on other complementary cases in which growth occurs either through concurring densification processes or through pure (local) isochoric transformations keeping $J_g=1$ that could be labeled as a morphogenic processes as well. Actually, prescribing the structure of the growth tensor reduces the role that the growth \textit{deviatoric} part --related as well-known to geometric distortions-- could have in describing internal material re-organization as a local driver of morphogenesis. %In this sense, a first point needing investigation would be therefore the role of deviatoric part of $\mathbf{F}_g$ and its relation to local morphogenesis. 
In the most of morphoelastic applications, changes in shape are provided for instance by considering differential volumetric growth between adjacent bodies in contact or as the result of growth-induced instability phenomena, which trigger a global configuration switch by exhibiting buckling, wrinkling or folding that re-organize the material towards stress-relaxing configurations \cite{ciarletta2012growth, klein2011experimental, amar2005growth, mirandola2023toward, goriely2005differential, huang2023mechanobiological}. These types of changes in form imply residual stress accumulation kindled by differential growth or constraints that cause the reconfiguration. In the same spirit, Ben Amar and co-workers simulate form changes of hyperelastic thin plates in \cite{dervaux2008morphogenesis, dervaux2009morphogenesis}, described by means of growth-induced curvature effects calculated directly from the growth strain. Therein, growth is interpreted as source of intrinsic curvatures, kindled by imposing anisotropic growth as well as inhomogeneous growth rates in nonlinearly growing plates. Similarly, Mahadevan and coworkers exploit differential growth to explain the mechanisms of morphogenesis of some slender structures in nature such as leaf and flowers \cite{liang2011growth} as well as of human organs, by analyzing the formation of folds in cortical brain \cite{tallinen2016growth}. They suggest that, in the so-called non-Euclidean elasticity, shaping can be viewed as an adaptation to inhomogeneous growth enforced by physical requirements, which select the actual shape of the structure by minimizing a certain elastic energy that takes origin from the incompatibility of intrinsic (natural) geometries and those induced by residual strains. In this respect, it has been recognized that although the the role of mechanics in driving growth is apparent and shaping emerges as a spontaneous consequence of metric ``frustration'' within a constrained/stressed environment, how actually the shape couples to local growth is an aspect of much current interest in biology \cite{lewicka2022geometry}. This concept is further explored by A. Goriely \cite{goriely2017mathematics}, who reports some still partly open challenges emrging from the efforts made for unifying the leading theoretical aspects of mechanobiology. In fact, although the above-mentioned examples set physically and geometrically consistent conditions for mechanistically simulating specific growth and morphogenic conditions, the precise form that the growth law should take is not completely understood as well as how the latter one could be actually put in direct relation to the complex evolution dynamics of a living system, which incorporates a vast class of processes that often involve coupled multiphysics descriptions with chemo-mechanical feedback across the scales of the several biological constituents. This would enhance morphoelasticity approaches by defining a strategy to investigate how morphogenic gradients, chemical pathways, physical inputs and mechanical transduction (more proper to the mechanobiology descriptions) combine to determine global size and shaping through local growth. In this regard, particular attention is put on determining the respective role of diffusion and mechanics in establishing biological patterns. In particular, identifying systems dominated by one or the other effect and fully-coupled systems is necessary to trace back development and generate patterning, this calling into play an inherent crosstalk with diffusion-reaction-driven approaches to morphogenesis such as the seminal problems carried out by Turing for describing phenomena of embryogenesis and phyllotaxis \cite{turing1990chemical}.
From all these observations, it emerges that a critical aspect the morphoelasticity problem \eqref{0prob} resides in the formal connection between geometry and mass balance. Mass conservation equation should be indeed ascribed as the sole effective constraint equation describing the dynamics of body's evolution by accounting for physically constitutive properties including diffusivity driving material flow as well as micro-architectural arrangements and species characteristic rates that govern the material development within a stressed environment, where all quantities can possibly scale, as well known, with stress or deformation in a measurable manner, and should be thus taken into proper account.
The morphoelasticity approach needs in fact to introduce constitutive relations for $\mathbf{F}_g$, or equivalent evolution laws for its growth rate $\dot{\mathbf{F}_g}$. However, the kinematics growth law somehow reduces the central role of the mass balance. This latter one is enforced to respect a constitutive equation for the sole volumetric growth \cite{ambrosi2017mechanobiology, lubarda2002mechanics, nappi2016stress, ambrosi2007stress}, but the overall growth kinematics instead evolves according to prescribed tensor relation independently on the natural directions potentially delineated by the mass rate. In this sense, specifying a kinematical description of the tensor growth law results a necessary trade-off to close \textit{ab initio} the morphoelastic problem, by overcoming the use of a single scalar equation. The growth law coupled to system \eqref{0prob} can be written in its general form as \cite{goriely2017mathematics, taber2020continuum, ciarletta2013mechano}:

\begin{equation}\label{Fgrate0}
    \mathbf{F}_g^{-1}\dot{\mathbf{F}}_g=\boldsymbol{\mathcal{G}}(\mathbf{P}-\mathbf{P}^*, c_i; \mathbf{X},t),
\end{equation}

where the general function $\boldsymbol{\mathcal{G}}$ often depends on mechanical and biochemical fields. Many ''stress-growth strain rate'' constitutive relations have been proposed in the literature for tracing back the growth/remodelling tensors, such as in the so-called homeostatic descriptions of growth where the stress $\mathbf{P}$ (or a related stress measure) is used as tracker for evolving towards prescribed environmental target conditions with homeostatic stress $\mathbf{P}^*$ applied to tumor progression, bone adaptation and arterial remodelling by considering particular specific mechanosensing functions and growth stimuli \cite{rodriguez1994stress, katsamba2020biomechanical, cyron2017growth, fraldi2018cells, ambrosi2017mechanobiology, ambrosi2007stress}. Other phenomenological kinematic laws of internal rearrangement inspire to crystal plasticity approaches and are driven by configurational forces \cite{ganghoffer2010eshelby, grillo2012growth, gurtin2010mechanics, ciarletta2012configurational}, recent models interpreting the growth as a plastic flow associated to the achievement of a homeostatic equivalent stress \cite{lamm2022macroscopic}. Interestingly, recent works focused on the active role of growth incompatibility in regulating size and generating residual stresses during morphogenesis \cite{erlich2024incompatibility, erlich2025geometric}, the mechanical and biochemical feedback at different scales establishing a possible connection between mass development and geometric growth being a further aspect to be included. The adopted evolution laws, although respecting thermodynamic restrictions, seem to suggest that --in absence of unique physical principles allowing for the identification of suitable forms of the growth law-- the morphoelastic problem still depends on \textit{a priori} assumed phenomenological descriptions of the growth tensor. This reduces its generality, the degree of freedom spent in postulating the growth structure potentially producing a paradoxical not uniqueness of the problem. 
In order to overcome this limitation, a remarkable strategy to generalize the morphoelasticity approach has been developed in Grillo et al. \cite{grillo2025approach}, in which the growth rate equation is re-interpreted as a nonholonomic constraint accounting for configurational forces and dissipative effects, so deriving the dynamics of mass, geometry, and material properties within a rigorous analytical mechanics framework. However, some chemically-based events due to possible independent evolution of the density field, which could enforce new unknowns in the growth problem with additional configurational terms and related stresses \cite{arricca2023coupled}, are not fully discussed. To deeply link mechanics and biology and contribute to generalize the morphoelasticity theory within a pure mechanobiological framework, a possible solution could be to find a more explicit connection between the master mass balance equation and the growth law. n this sense, the mass balance cannot only partially relate to the volumetric part of the kinematics, but the overall structure of the $\mathbf{F}_g$ rate --i.e. of both its volumetric and deviatoric part-- should naturally emerge from mass evolution by exploiting the knowledge of the physical micro-structural features of the material, whose rearrangement drive the transmission of mechanical stresses and strain-based stimuli.
According to this hypothesis, the mass balance would \textit{de facto} constitute the sole effective regulator of growth kinematics not only in terms of both volumetric expansion and density change but also in terms of local morphogenesis. In this sense, the unique physical principle already resides in the morphoelastic problem but it must be opportunely generalized by considering the knowledge of all the measurable physical properties of the material. 
In order to find this type of relation and delve into the different points above discussed, we start from the well-established morphoelastic framework to propose a modelling strategy in which the generalization of the mass balance leads to close the growth problem, the inherent rate of the materoal being put in a one-to-one connection with the evolution of geometry and density so as to describing growth, remodelling and morphogenesis phenomena. In fact, besides returning the classical assumptions made for the volumetric growth, the presented procedure can be of crucial interest to further assess the way in which mass balance can steer also the geometric distortions linked to the deviatoric part of the growth. These represent by definition local morphogenetic transformation of the body in its natural state, obtained through following material fluxes and rearrangement of the internal micro-architectures, and so avoiding the mechanical interplays needed in classical morphoelastic cases.
It is felt that considering a biomechanically consistent problem in which the kinematic growth law is entirely resolved from the respect of the mass balance, so as to bridge in a natural manner diffusion and mechanics, can represent a novel direction to fully meld the information from systems' biology with the overall deformation, stresses and evolving constitutive properties of a living tissue, in this way helping to take a further step towards an unified view of the leading mechanobiology processes.

\section{Paradoxical not uniqueness of canonical growth laws}

The kinematics of a growing body can be described by starting from the canonical multiplicative decomposition of the deformation gradient $\mathbf{F}=\mathbf{I}+\mathbf{u}\otimes\nabla_\mathbf{X}$ --the vector $\mathbf{u}$ representing the displacement field of material points $\mathbf{X}$-- into a growth part and an elastic part, i.e. $\mathbf{F}=\mathbf{F}_e\,\mathbf{F}_g$ \cite{rodriguez1994stress}. In this view, each material point with mass $dm_0=\rho_0\,dV_0$ belonging to a reference configuration $\mathbf{X}\in\mathcal{B}_0$ is first mapped onto a generally incompatible grown configuration $\mathcal{B}_g$, where the element gains mass $dm_g=\rho_g\,dV_g=\rho_g\,J_g\,dV_0$. Then, mass element is mapped on a current configuration $\mathcal{B}$, in which elementary constituents deform elastically to internally adapt each other and to respond to the eventual presence of applied loads, their updated mass and position being here expressed respectively as $dm=\rho\,dv$ and $\mathbf{x}\in\mathcal{B}$, with $\mathbf{x}=\mathbf{X}+\mathbf{u}$ and $dv=J\,dV_0$. Without loss of generality, the balance of mass that transforms body elements with mass $\rho_0\,dV_0$ to body elements with mass $\rho\,dv$ reads as \cite{goriely2017mathematics, gurtin2010mechanics}:

\begin{equation}\label{mb1}
\int dm= \int dm_0 - \int_0^t \int_{\partial v}\mathbf{q}\cdot \mathbf{n}\, ds + \int_0^t \int_{v} \gamma \rho dv
\end{equation}

where the right hand of the equation reports in explicit both possible species diffusion driven by a flux vector $\mathbf{q}$ and a source/sink term $\gamma$ representing a general growth rate that, in the most general formulations, can include the influence of mutually interacting species as well as \textit{ad hoc} modelled mechano-chemical feedback mechanisms \cite{fraldi2018cells,lorenzo2016tissue, ackermann2025mechanistic, bernard2024modelling}. Taking into account that mass does not vary in elastic adaptation, it results  $dm=dm_g$. This implies that the balance of mass \eqref{mb1} can be rewritten after some passages with respect to the natural configuration in the usual rate form \cite{jones2012modeling}:

\begin{equation}\label{mb2}
\dot{\overline{\rho_g\, J_g}}=-\nabla_0\cdot \mathbf{Q}+\rho_g\, J_g \Gamma=\rho_g\, J_g\left[-\frac{1}{\rho_g\, J_g}\nabla_0\cdot \mathbf{Q}+\Gamma\right]=\rho_g\, J_g\, \mathcal{R}_g
\end{equation}

or 
\begin{equation}\label{mb3}
\frac{\dot{\rho}_g}{\rho_g}+\frac{\dot{J}_g}{J_g}=\frac{\dot{\rho}_g}{\rho_g}+tr(\dot{\mathbf{F}}_g\mathbf{F}_g^{-1})=\mathcal{R}_g
\end{equation}

where the material flux vector $\mathbf{Q}$ and rate $\Gamma$ have been introduced. Equation \eqref{mb3} shows that the net between mass transport and generation produces combined density and/or volume changes. When referring to a biological material, these growth phenomena could be associated to different concurring processes taking place at the microscale level, such as cell proliferation and migration within tissue interstitial spaces as well as ECM synthesis and degradation, with effect on the body material properties and microstructural reorganization. However, in this form, the scalar mass balance \eqref{mb3} does not provide any hypothesis either on how mass accumulation/removal affects density and volume separately or on how mass kinematics distributes in space along the different directions. In the most of applications, a purely volumetric growth process is assumed \cite{lubarda2002mechanics, nappi2016stress}, the density changes are entirely elastic effects so that the following relations hold true: 

\begin{equation}\label{mb4}
\rho_g=\rho_0, \quad \rho=\rho_0/J_e \quad \text{and} \quad  \frac{\dot{J}_g}{J_g}=\mathcal{R}_g
\end{equation}

In this way, the source/sink of mass coherently drives volumetric growth by providing a single scalar equation that gives information only about the determinant of the growth tensor. This would imply that, for a given evolution law, different growth tensors with a shared determinant could be in principle obtained. The last equation \eqref{mb4} indeed describes \textit{bulk growth} by not specifying how growth spatially distributes within the body. A possible strategy provides the partition of the volumetric growth in the three material directions through prescribed anisotropy factors, eventually depending on mechanical arguments \cite{katsamba2020biomechanical, kuhl2014growing, arumugam2019model}. If on a side one has the volumetric part of the growth is fully characterized by the mass equation, it seems that additional information is needed for the remaining isochoric part that, basically related to shape distortions. This would make the growth problem somehow incomplete, by requiring to accompany the constitutive relation \eqref{mb4} with a phenomenological tensor equation of the type $\dot{\mathbf{F}}_g\mathbf{F}_g^{-1}=\boldsymbol{\mathcal{G}}$ \cite{goriely2017mathematics, taber2020continuum}, where the general function $\boldsymbol{\mathcal{G}}$ involves all the needed couplings with mechanical and biochemical fields. However, such a kinematical relation is not uniquely identified in the most of cases and no direct connection is established with the mass balance, which should represent the sole effective constraint of the growth problem. Then, starting from the widely consolidated framework of volumetric growth, we here wonder about what is intended with anisotropic growth and what could be the role of the growth deviatoric part in driving morphogenesis. With this motivation, we seek for a natural way to identify the isochoric part of the growth in the respect of the mass balance and starting from the relevant material features, which provide as data the knowledge of tissue-specific architectures involved in the growth process at different scale levels. 
In order to explore this intriguing paradox, here we propose a straightforward manner to uniquely identify the whole growth tensor by exploiting the mass balance and by suggesting a novel description of some local mechanisms concurring to macroscopic morphogenesis of living structures passing through the deviatoric part of the growth tensor.

\section{A heuristic approach to close the growth mechanics problem}
\subsection{Generalization of the mass balance and kinematics of growth, remodeling and morphogenesis}\label{sec:genmass}

In order to highlight the connection of mass balance with intrinsic material anisotropy and change of shape, we consider a vector description of the geometric and mass distribution within the body element. In this sense, standard approaches let us interpret the generic mass element in a given configuration as the Jacobian of elements with linear mass, which si distribute the mass along the three dimensions of the volume element in a possibly anisotropic manner. This means that, within each element, a density anisotropy tensor has to be considered to take into account these differences. In the natural  configuration, the mass vector reads:

\begin{equation}\label{dmg}
    dm_g=\det [d\mathbf{m}_{g,i}], \quad d\mathbf{m}_{g,i}=\rho_g^{1/3}\,\boldsymbol{\Upsilon}_g\cdot d\mathbf{X}_{g,i}= \rho_g^{1/3}\,\boldsymbol{\Upsilon}_g\cdot\mathbf{F}_g\cdot d\mathbf{X}_{0,i}
\end{equation}

where $\boldsymbol{\Upsilon}_g$ is a second rank tensor formally introduced to account for density anisotropy and defined such that $\det\boldsymbol{\Upsilon}_g=1$, tracing back the evolution of material directions. Without loss of generality, the density-associated tensor can be assumed as intrinsically symmetric. Differently, the growth tensor is generally not symmetric and it is assumed to owe strictly positive eigenvalues to ensure positive mass and volume changes, although this hypothesis does not generally pass through the introduction of a proper growth deformation metrics \cite{goriely2017mathematics}. In order to distinguish the contribution of bulk growth and the \textit{morphic} term, let us split the growth tensor into its volumetric and isochoric parts, i.e. 

\begin{equation}\label{Fgsplit}
\mathbf{F}_g=J_g^{\frac{1}{3}}\,\mathbf{M}, \quad \det\mathbf{M}=1
\end{equation}

Herein, the isochoric part is responsible of natural geometric distortions and change of shape during the growth process and could be thus labeled as a local inelastic regulator of morphogenesis. Although the volumetric part of the growth tensor is straightforwardly related to mass balance, this morphogenic term lacks of a direct identification with the associated evolution problem. In view of \eqref{Fgsplit}, equation \eqref{dmg} can be rewritten as

\begin{equation}\label{dmg2}
    d\mathbf{m}_{g,i}= \rho_g^{1/3}\,J_g^{1/3}\,\boldsymbol{\Upsilon}_g\cdot\mathbf{M}\cdot d\mathbf{X}_{0,i} =  \boldsymbol{\rho}_g\cdot d\mathbf{X}_{0,i}, \quad \det \boldsymbol{\rho}_g= \rho_g\, J_g
\end{equation}

This equation shows that the evolution of the mass with respect to the reference directions is entirely governed by the rate of the condensed operator $\boldsymbol{\rho}_g$, which includes both kinetics and density effects. In this sense, equation \eqref{dmg2} let to generalize the rate equation \eqref{mb2} in view of the Jacobi formula:

\begin{equation}\label{genrate1}    
    \frac{\dot{\overline{\rho_g\, J_g}}}{\rho_g\, J_g}=tr(\boldsymbol{\rho}_g^{-1}\,\dot{\boldsymbol{\rho}}_g)= \mathcal{R}_g = -\frac{1}{\rho_g\, J_g}(\nabla_0\cdot \mathbf{Q})+\Gamma
\end{equation}

This allows us to assume that the net growth rate $\mathcal{R}_g$ can be seen as the trace of a more generic evolution tensor, which distributes fluxes and generation along the reference directions. In particular, we have (see \hyperref[AppA]{Appendix} for details):

\begin{equation}\label{genRg}
    \mathcal{R}_g= \mathbf{I}:\boldsymbol{\mathcal{R}}_g, \quad \boldsymbol{\mathcal{R}}_g=\mathbf{F}_g^{-1}\,\boldsymbol{r}_g\mathbf{F}_g=-\frac{1}{\rho_g\, J_g}\left(\mathbf{Q}\otimes \nabla_0\right)+\mathbf{M}^{-1}\,\mathbf{H}\,\mathbf{M}\, \Gamma
\end{equation}

Herein, the material growth rate is given in explicit by two contributions, one related to the gradient of the flux $\mathbf{Q}$ and tracing the capability by which species can redistribute mass elements through diffusion. In this form, the flux $\mathbf{Q}$ also accounts for possible not accumulative (solenoidal) fluxes that redistribute mass in a convective manner within the volume. In addition, equation \eqref{genRg} is generalized by introducing a second term constituted by the intrinsic rate weighted by the material tensor $\mathbf{M}^{-1}\,\mathbf{H}\,\mathbf{M}$, which is by definition normalized such that $tr(\mathbf{M}^{-1}\,\mathbf{H}\,\mathbf{M})=tr(\mathbf{H})=1$. More specifically, the tensor $\mathbf{H}$ represents a second order fabric tensor introduced to take into account for the natural micro-structural organization of the grown material, which might present peculiar architecture able to steer the growth throughout the different directions. This fabric tensor represents an additional structural information of the material that is not \textit{a priori} postulated but it can be constructed directly by exploiting well-established techniques in order to highlight tissue-specific arrangement at each material point \cite{cowin1989identification, jemiolo1997fabric, moreno2014techniques, moreno2014techniques, marques2018multiscale, tabor2009equivalence}. 
In the light of these positions, different sources of anisotropy can be identified, one related to the permeability governing mass walkway and the other one associated to material distribution, which directly affects how growth orients in space as well as the elastic anisotropy of the material, with consequent impact on its mechanical behavior and stress re-distribution \cite{turner1987dependence, hariton2007stress}. It is worth to notice that, in general, the scaffold and the elastic structure of the tissue do not present the same degree of anisotropy, since the architectural networks inviting proliferation and movement can be different from the ones describing tissue mechanical response. By the way of example, there could be cases in which, in absence of preferential formations that polarize the growth such as oriented vascular networks, mass can grow isotropically also within a material exhibiting constitutive (elastic) anisotropies. On the other hand, especially with reference to large deformations \cite{carotenuto2019growth}, growth can concur to the material remodelling of the elastic constants of tissue by determining residual stress-induced inhomogeneities and anisotropies. In such a case, possible growth anisotropies can modify the elastic behaviour of the tissue, and the fabric could affect the elastic symmetry of the virgin material \cite{cowin1992evolutionary}. In the light of position \eqref{genRg}, balance \eqref{genrate1} can be generalized as follows

\begin{equation}\label{genrate2}    
    \boldsymbol{\rho}_g^{-1}\cdot\dot{\boldsymbol{\rho}}_g=\frac{1}{3}\frac{\dot{\rho}_g}{{\rho}_g}\mathbf{I} +\frac{1}{3}\frac{\dot{J}_g}{{J}_g}\mathbf{I} + \mathbf{M}^{-1}\cdot\overset{\circ}{\mathbf{L}}_Y\cdot \mathbf{M}+\overset{\circ}{\mathbf{L}}_M=\boldsymbol{\mathcal{R}}_g = \frac{1}{3}\,\mathcal{R}_g\mathbf{I}+\boldsymbol{\mathcal{R}}_g^{dev}
\end{equation}
%(\mathbb{I}+\mathbb{E}_{\textbackslash}):
%$\mathbb{I}$ and $\mathbb{E}_{\textbackslash}$ are respectively the fourth order identity tensor and a fourth order operator respecting the condition $\mathbf{I}:\mathbb{E}_{\textbackslash}=0$ and mediating the possible combination of mass rate terms along the different material directions, which thus affects
%Since this generalization may be not unique, the expression of the latter term has to be properly defined in the following.
%On the other hand, $\mathbb{I}$ and $\mathbb{E}_{\textbackslash}$ respectively represent the fourth order identity tensor and a fourth order operator respecting the condition $\mathbf{I}:\mathbb{E}_{\textbackslash}=\textbf{0}$ and mediating the possible combination of mass rate terms along different material directions, whose expression is introduced in the following.

in which $\overset{\circ}{\mathbf{L}}_Y=\boldsymbol{\Upsilon}_g^{-1}\dot{\boldsymbol{\Upsilon}}_g$ and $\overset{\circ}{\mathbf{L}}_M=\mathbf{M}^{-1}\dot{\mathbf{M}}$ are respectively the material rates of density anisotropy and morphogenesis, while the deviatoric term of mass rate is defined such that $\boldsymbol{\mathcal{R}}_g^{dev}=[\mathbb{I}-(1/3)(\mathbf{I}\otimes\mathbf{I})]:\boldsymbol{\mathcal{R}}_g=\mathbb{P}^{dev}:\boldsymbol{\mathcal{R}}_g$. By direct identification, one can split the problem \eqref{genrate2} in the classical mass balance plus a deviatoric rest, i.e.

\begin{equation}\label{sys1}
    \begin{cases}
        &\frac{\dot{\rho}_g}{{\rho}_g}+\frac{\dot{J}_g}{{J}_g}=\mathcal{R}_g\\
        &\mathbf{M}^{-1}\cdot\overset{\circ}{\mathbf{L}}_Y\cdot \mathbf{M}+\overset{\circ}{\mathbf{L}}_M=\boldsymbol{\mathcal{R}}_g^{dev}
    \end{cases}
\end{equation}

The second equation describes how mass evolution, by growing along with fabric directions and by moving material driven by intrinsic transport mechanisms, is capable to regulate density re-organization and geometric distortion. Importantly, this microstructural rearrangement is mediated by diffusivity and fabric tensors, which can be derived through standard micromechanical considerations. It is reasonable to assume a strict connection between the density anisotropy tensor ${\boldsymbol{\Upsilon}}_g$ and the adopted measure of fabric tensor $\mathbf{H}$ in equation \eqref{genRg}, since both account for the anisotropy generated by the internal material distribution. Also, since microstructure evolves according to density and geometric changes, the respective rates should be somehow put in direct relation each other. As shown in the \hyperref[AppA]{Appendix}, by starting from well-established arguments in the characterization of material micro-architectures, a selected measure of the fabric tensor $\mathbf{H}$ can be seen as correlated to both density and morphogenesis rates. In particular, it is shown that fabric rate can be put in the following form:
%%QUI_NOW

\begin{align}\label{Hdot}
    &\mathbf{H}=\tilde{\mathbf{H}}(\rho_g,\nabla_0\rho_g, \mathbf{M})=\frac{1}{3}\mathbf{I}+\mathbf{K},\nonumber\\
    &\dot{\mathbf{H}}=\dot{\mathbf{K}}=\frac{1}{3}\mathbb{\Omega}_M^{dev}:\overset{\circ}{\mathbf{L}}_M+\frac{1}{3}\mathbf{\Omega}_\rho^{dev}\,\frac{\dot{\rho}_g}{{\rho}_g}, \quad \mathbf{I}:\dot{\mathbf{H}}=0
\end{align}

where $\mathbf{\Omega}_\rho^{dev}$ and $\mathbb{\Omega}_M^{dev}$ are respectively a second and fourth-rank tensor that vanish contracted by the second order identity, and collecting rate functions that may depend on geometric growth deformation as well as by the density and density gradients. These terms result proportional to a characteristic internal length-scale $\ell_0^2$ along with the density-induced anisotropy is measured (see \hyperref[AppA]{Appendix} for details). On the other hand, in the present description, ${\boldsymbol{\Upsilon}}_g$ represents a meso-scale description of density anisotropy in the grown/densified configuration. Therefore, starting from the identity

\begin{equation}\label{trHY}
    \mathbf{I}:\boldsymbol{\Upsilon}_g^{-1}\dot{\boldsymbol{\Upsilon}}_g=\mathbf{I}:\dot{\mathbf{K}}=0
\end{equation}

We then make the \textit{ansatz} that the two anisotropy descriptors are strictly correlated between each other, the rate of density rearrangement tensor resulting \textit{de facto} proportional to the incremental change of the material fabric $\dot{\mathbf{H}}$. This means to establish a relation of the type $\overset{\circ}{\mathbf{L}}_Y=\mathbf{f}(\dot{\mathbf{K}})$, which depends on the measure by which the fabric of the material is computed. As shown in detail in the \hyperref[AppA]{Appendix}, for the adopted fabric definition, the following relation can in particular established: 

\begin{align}\label{trHY2}
&\overset{\circ}{\mathbf{L}}_Y \simeq -3\dot{\mathbf{K}}= -\mathbb{\Omega}_M^{dev}:\overset{\circ}{\mathbf{L}}_M-\mathbf{\Omega}_\rho^{dev}\,\frac{\dot{\rho}_g}{{\rho}_g}
\end{align}

This rate equation actually identifies the fabric tensor as a first-order approximation of the nonlinear measure ${\boldsymbol{\Upsilon}}_g$ (see \hyperref[AppA]{Appendix}), the symmetry hypothesis here also implying that the difference between the spin in the geometrically grown and the (pulled back) spin in the grown-and-densified configuration is actually negligible (see e.g. \cite{lubarda2001elastoplasticity}). In light of equation \eqref{trHY2}, equation \eqref{sys1}$_2$ becomes

\begin{equation}\label{LMeq}
    \mathbb{V}_M:\overset{\circ}{\mathbf{L}}_M=\boldsymbol{\mathcal{R}}_g^{dev} + \mathbf{M}^{-1}\,\mathbf{\Omega}_\rho^{dev}\,\mathbf{M}\,\frac{\dot{\rho}_g}{{\rho}_g},
    \qquad \mathbb{V}_M=\mathbb{I}-\left(\mathbf{M}^{-1}\underline{\overline{\otimes}}\,\mathbf{M}^T\right):\mathbb{\Omega}_M^{dev}
\end{equation}

%We then introduce as \textit{ansatz} the possibility that devitoric geometric changes and mass rate terms evolve coaxially by posing $\mathbb{W}_M=(\mathbb{I}+\mathbb{E}_{\textbackslash})$, 

It therefore follows that

\begin{align}\label{LMeq2}
    \overset{\circ}{\mathbf{L}}_M &=\mathbb{\Lambda}_M:\left[\boldsymbol{\mathcal{R}}_g^{dev} + \mathbf{M}^{-1}\,\mathbf{\Omega}_\rho^{dev}\,\mathbf{M}\,\frac{\dot{\rho}_g}{{\rho}_g}\right], \nonumber\\
    &\mathbb{\Lambda}_M=\mathbb{V}_M^{-1}\simeq \mathbb{I}+\left(\mathbf{M}^{-1}\underline{\overline{\otimes}}\,\mathbf{M}^T\right):\mathbb{\Omega}_M^{dev}
\end{align}

where the inverse matrix $\mathbb{\Lambda}_M$ is estimated on the basis of the considered small sub-macroscopic lengths weighting density gradients (see \hyperref[AppA]{Appendix}). System \eqref{sys1} then accordingly modifies by giving direct relations to identify the evolution of the whole growth tensor by tracing back both morphogenesis and remodelling kinematic terms

\begin{equation}\label{sys2}
    \begin{cases}
        &\frac{\dot{\rho}_g}{{\rho}_g}+\frac{\dot{J}_g}{{J}_g}=\mathcal{R}_g\\
        &\overset{\circ}{\mathbf{L}}_M=\mathbb{\Lambda}_M:\left[\boldsymbol{\mathcal{R}}_g^{dev} + \mathbf{M}^{-1}\,\mathbf{\Omega}_\rho^{dev}\,\mathbf{M}\,\frac{\dot{\rho}_g}{{\rho}_g}\right]
 %       &\mathbf{M}^{-1}\cdot\overset{\circ}{\mathbf{L}}_Y\cdot \mathbf{M}=\mathbb{\Omega}_M^{dev}:\overset{\circ}{\mathbf{L}}_M
    \end{cases}
\end{equation}

%In this way, it is possible to assume that fluxes and directional growth invited by the micro-architecture redistribute the mass within the transforming volume so as to induce a new material disposition, in which density reorganization can be indirectly traced back from combined density evolution and geometric distortions. Under these assumptions, we here assume that morphogenesis velocity terms can be one-by-one identified with mass rate deviatoric terms by setting:
%\begin{equation}
%   \mathbb{\Omega}_M^{dev}=\mathbb{E}_{\textbackslash}\quad \text{and} \quad \overset{\circ}{\mathbf{L}}_M=\boldsymbol{\mathcal{R}}_g^{dev}
%\end{equation}
%\textcolor{red}{It is worth to note that the operator $\mathbb{\Lambda}_M$ depends on the sub-macroscopic length scale involved in the description of local density stereological distribution, and in situation of practical interest in which this characteristic length is much smaller than the size of the macroscopic representative volume element one has that $\mathbb{\Lambda}_M\simeq\mathbb{I}$, so achieving a one-to-one correspondence between the morphogenetic inelastic deformation and deviatoric mass rates.}

According to this system, mass rate causes growth and material evolution by determining new grown and morphed material vectors as well as density evolution in the natural configuration, which together shape the material architecture by updating fabric and material (density) anisotropies in a way proportional to the deviatoric part of mass production. More importantly, these evolution laws are all defined from the knowledge of he mass balance, by so completely connecting the well-established kinematics of growth with the underlying mechanobiology of the considered living system through a heuristic process that avoids the need of postulating structures of either the growth tensor or its rate. Also, this approach takes advantage of all the constitutive and microstructural elements that characterize the material at a macroscopic scale, with the possibility of following how they dynamically evolve as a function of growth, remodelling and morphogenesis. 
Coupling \eqref{sys2} to \eqref{0prob}, the growth problem re-writes as 

\begin{equation}\label{0prob2}
    \begin{cases}
        &\nabla_0\cdot\mathbf{P}=-\textbf{B}\\
        &\frac{\dot{\rho}_g}{{\rho}_g}+\frac{\dot{J}_g}{{J}_g}=\mathcal{R}_g\\
        &\overset{\circ}{\mathbf{L}}_M=\mathbb{\Lambda}_M:\left[\boldsymbol{\mathcal{R}}_g^{dev} + \mathbf{M}^{-1}\,\mathbf{\Omega}_\rho^{dev}\,\mathbf{M}\,\frac{\dot{\rho}_g}{{\rho}_g}\right]
     %   &\overset{\circ}{\mathbf{L}}_Y=\mathbf{M}\,\mathbf{\Omega}_\rho^{dev}\,\mathbf{M}^{-1}\,\frac{\dot{\rho}_g}{{\rho}_g}
 %       &\mathbf{M}^{-1}\cdot\overset{\circ}{\mathbf{L}}_Y\cdot \mathbf{M}=\mathbb{\Omega}_M^{dev}:\overset{\circ}{\mathbf{L}}_M
    \end{cases}
\end{equation}

that constitutes a system of $12$ equations in the $13$ unknowns $\{\rho_g,J_g,\mathbf{u},\mathbf{M}\}$ (the condition $\det \mathbf{M}=1$ reduces by 1 the independent components of $\mathbf{M}$). This implies that one additional condition is needed to properly close the problem, which can be sought after from studying the constitutive framework.

\section{Constitutive framework}

The constitutive assumptions for the kinematic variables of equation \eqref{sys2} can be then derived from the combined energy-entropy inequality in the well-known Colemann and Noll's sense. In the case of the general chemo-mechanical problem at hand, this includes both mechanical power terms as well as entropic terms due to mass fluxes and internal generation, which are directly conjugated to growth- and remodelling-induced microstructural changes, defined per unit reference volume \footnote{A generic driving force per unit mass is expressed per unit reference volume as $\int_m\,[\mathcal{A}]\,dm = \int_{V_0}\,J_g\,(\rho_g/\rho_0)[\mathcal{A}_0]\,dV_0$} \cite{nappi2016stress, lubarda2002mechanics, fraldi2018cells}. By introducing a free energy density $\psi$, the Clausius-Duhem inequality reads as:

 \begin{footnotesize}
\begin{align}\label{diss1}
    &\int_{V_0} \mathbf{S}:\frac{1}{2}\dot{\mathbf{C}}\,dV_0 - \int_{S_0}\mathbf{N}\cdot\left(\frac{\boldsymbol{\mathcal{X}}}{\rho_0}\right)\cdot\mathbf{Q}\,dS_0 + 
    \int_{V_0} J_g\frac{\rho_g}{\rho_0}(\boldsymbol{\mathcal{X}}:\mathbf{M}^{-1}\,\mathbf{H}\,\mathbf{M})\,\Gamma\,dV_0 + \int_{V_0} J_g\frac{\rho_g}{\rho_0}\,\psi\,\mathcal{R}_g\,dV_0 \geq \nonumber\\
    & \geq \frac{d}{dt}\int_{V_0} J_g\frac{\rho_g}{\rho_0}\,\psi(\mathbf{C}_e, \frac{\rho_g}{\rho_0}, \mathbf{H}, \nabla_0\times\mathbf{F}_g) \,dV_0 +\int_{V_0} \mathbf{Q}\cdot \boldsymbol{\kappa}^{-1}\cdot\mathbf{Q}\,dV_0
\end{align}
\end{footnotesize}

where $\mathbf{C}=\mathbf{F}^T\mathbf{F}$ and $\mathbf{C}_e=\mathbf{F}_e^T\mathbf{F}_e$ are Green-type (total and elastic) strain tensors, $\mathbf{S}=\mathbf{F}^{-1}\mathbf{P}$ is the 2-nd Piola Kirchhoff stress tensor. On the other hand, $\boldsymbol{\mathcal{X}}$ represents a generalized chemo-mechanical driving force (defined as a microscopic reference stress, see e.g. \cite{gurtin2010mechanics}) that is power-conjugate to local growth and morphogenesis. Also, the last term on the left-hand side of the equation represents the power contribution of chemo-mechanical energy per unit mass production \cite{lubarda2002mechanics}. On the right-hand side of \eqref{diss1}, the free-energy is here assumed to depend respectively on the elastic strain measure $\mathbf{C}_e$, on the density change ${\rho_g}/{\rho_0}$ as well as on possible microstructural (mesoscale) contributions related to the anisotropic fabric $\mathbf{H}$ and the growth-asociated dislocation density, i.e. $\nabla_0\times\mathbf{F}_g$ (see e.g. \cite{kaiser2019incompatibility, kaiser2019dislocation}). Additionally, the last term models the dissipation due to species transport mediated by a friction coefficient matrix. Under these positions, one can write 

\begin{footnotesize}
 \begin{align}\label{diss2}
    &\int_{V_0}\quad \mathbf{S}:\frac{1}{2}\dot{\mathbf{C}} -  \mathbf{Q}\cdot \left(\frac{\boldsymbol{\mathcal{X}}}{\rho_0}\right)\cdot \nabla_0  + J_g\frac{\rho_g}{\rho_0}\boldsymbol{\mathcal{X}}:\left[-\frac{1}{J_g\rho_g}(\mathbf{Q}\otimes\nabla_0)+(\mathbf{M}^{-1}\,\mathbf{H}\,\mathbf{M})\Gamma \right] + J_g\frac{\rho_g}{\rho_0}\,\psi\,\mathcal{R}_g \quad dV^0 \geq\nonumber\\
    &\int_{V_0}\quad\frac{\dot{\overline{J_g\rho_g}}}{\rho_0} \psi + 
    J_g\frac{\rho_g}{\rho_0}\left[\frac{\partial \psi}{\partial \mathbf{C}_e}:\dot{\mathbf{C}}_e + \frac{\rho_g}{\rho_0}\frac{\partial \psi}{\partial (\rho_g/\rho_0)}\frac{\dot{\rho_g}}{\rho_g}+\frac{\partial \psi}{\partial \mathbf{H}}:\dot{\mathbf{H}}+\frac{\partial \psi}{\partial \boldsymbol{\vartheta}}:  (\nabla_0\times\dot{\mathbf{F}}_g)\right]+ \mathbf{Q}\cdot \boldsymbol{\kappa}^{-1}\cdot\mathbf{Q} \quad dV^0
\end{align}
\end{footnotesize}

where $\boldsymbol{\vartheta}=\nabla_0\times \mathbf{F}_g$. On the right-hand side, by exploiting the identity $\mathbf{A}:(\nabla_0\times\mathbf{B})=\nabla_0\times(\mathbf{A}\mathbf{B}):\mathbf{I} + \nabla_0\times\mathbf{A}:\mathbf{B}$, one has 

\begin{footnotesize}
 \begin{align}\label{diss2b}
    &\int_{V_0}\quad \mathbf{S}:\frac{1}{2}\dot{\mathbf{C}} -  \mathbf{Q}\cdot \left(\frac{\boldsymbol{\mathcal{X}}}{\rho_0}\right)\cdot \nabla_0  + J_g\frac{\rho_g}{\rho_0}\boldsymbol{\mathcal{X}}:\left[-\frac{1}{J_g\rho_g}(\mathbf{Q}\otimes\nabla_0)+(\mathbf{M}^{-1}\,\mathbf{H}\,\mathbf{M})\Gamma \right] + J_g\frac{\rho_g}{\rho_0}\,\psi\,\mathcal{R}_g \quad dV^0 \geq\nonumber\\
    &\int_{V_0}\quad\frac{\dot{\overline{J_g\rho_g}}}{\rho_0} \psi + 
    J_g\frac{\rho_g}{\rho_0}\left[\frac{\partial \psi}{\partial \mathbf{C}_e}:\dot{\mathbf{C}}_e + \frac{\rho_g}{\rho_0}\frac{\partial \psi}{\partial (\rho_g/\rho_0)}\frac{\dot{\rho_g}}{\rho_g}+\frac{\partial \psi}{\partial \mathbf{H}}:\dot{\mathbf{H}}+ \frac{\rho_0}{\rho_g J_g}\mathbf{F}_g^T\,(\nabla_0\times\mathbf{\Theta}):  \mathbf{F}_g^{-1}\dot{\mathbf{F}}_g\right]+ \mathbf{Q}\cdot \boldsymbol{\kappa}^{-1}\cdot\mathbf{Q} \quad dV^0
\end{align}
\end{footnotesize}

 in which $\mathbf{\Theta} =\partial\psi/\partial(\nabla_0\times\mathbf{F}_g)$ and the boundary condition $\mathbb{e}:\mathbf{\Theta}\cdot\mathbf{N}=\mathbf{0}$ is implied, $\mathbb{e}$ representing the Levi-Civita permutation third rank tensor \cite{kaiser2019incompatibility}. The selected state variables of the thermodynamic process are represented by the elastic strains, the density and the internal growth, i.e. $\{\mathbf{C}_e,\rho_g, \mathbf{F}_g\}$. Therefore, after opportune localization, the second and the last terms of dissipation \eqref{diss2b} allow one to first assume the fluxes proportional to the imbalance of chemo-mechanical driving forces through the relation

\begin{equation}\label{fluxass}
    \mathbf{Q}=- \boldsymbol{\kappa}\cdot\left(\frac{\boldsymbol{\mathcal{X}}}{\rho_0}\right)\cdot \nabla_0
\end{equation}

Furthermore, with the help of Eqs. \eqref{genRg}--\eqref{Hdot}, dissipation \eqref{diss2} can be reformulated as

\begin{align}\label{diss3}
    &\mathbf{S}:\frac{1}{2}\dot{\mathbf{C}} + J_g\frac{\rho_g}{\rho_0}\boldsymbol{\mathcal{X}}:\left[ \frac{1}{3}\mathcal{R}_g\mathbf{I}+\boldsymbol{\mathcal{R}}_g^{dev}\right]  \geq\nonumber\\
    &J_g\frac{\rho_g}{\rho_0}\left[\frac{\partial \psi}{\partial \mathbf{C}_e}:\dot{\mathbf{C}}_e + \frac{\rho_g}{\rho_0}\frac{\partial \psi}{\partial (\rho_g/\rho_0)}\frac{\dot{\rho_g}}{\rho_g}+\frac{\partial \psi}{\partial \mathbf{H}}:\left(\mathbf{\Omega}_\rho^{dev}\,\frac{\dot{\rho}_g}{{\rho}_g}+\mathbb{\Omega}_M^{dev}:\overset{\circ}{\mathbf{L}}_M\right) + \mathbf{F}_g^T\,(\nabla_0\times\mathbf{\Theta}):  \overset{\circ}{\mathbf{L}}_g \right]
\end{align}

In order to deal with well identified thermodynamical forces, let us express the effective chemo-mechanical stress as $\boldsymbol{\mathcal{X}}=\mu\,\mathbf{I}+\boldsymbol{\Sigma}:\mathbb{\Lambda}_M$, the term $\mu$ indicating the chemical potential and $\boldsymbol{\Sigma}$ the configurational stress. Under this position, and considering equations \eqref{sys2}$_2$, the dissipation becomes
%pulled-back from to grown-and-densified configuration to the grown reference configuration, on which elastic stretches are also defined

\begin{footnotesize}
\begin{align}\label{diss4}
    &\mathbf{S}:\frac{1}{2}\dot{\mathbf{C}} 
    + \underbrace{J_g\frac{\rho_g}{\rho_0}\,\mu\,\frac{\dot{\rho}_g}{{\rho}_g}}
    + J_g\frac{\rho_g}{\rho_0}\mu\,\frac{\dot{J}_g}{J_g}+ J_g\frac{\rho_g}{\rho_0}\left[\frac{1}{3}\boldsymbol{\Sigma}:\mathbf{I}\right]\frac{\dot{\rho}_g}{{\rho}_g}+
    %%%%
    \underbrace{J_g\frac{\rho_g}{\rho_0}\left(-\boldsymbol{\Sigma}:\mathbb{\Lambda}_M:\mathbf{M}^{-1}\mathbf{\Omega}_\rho^{dev}\mathbf{M}\right)\frac{\dot{\rho}_g}{{\rho}_g}}+ \\ \nonumber 
    %%%%
    &+ J_g\frac{\rho_g}{\rho_0}\boldsymbol{\Sigma}:\overset{\circ}{\mathbf{L}}_g  
    \geq\nonumber\\
    &J_g\frac{\rho_g}{\rho_0}\frac{\partial \psi}{\partial \mathbf{C}_e}:\dot{\mathbf{C}}_e
    +
    \underbrace{J_g\frac{\rho_g}{\rho_0}\left[\frac{\rho_g}{\rho_0}\frac{\partial \psi}{\partial (\rho_g/\rho_0)}+ \frac{1}{3}\frac{\partial \psi}{\partial \mathbf{H}}:\mathbf{\Omega}_\rho^{dev}\right]\frac{\dot{\rho_g}}{\rho_g}} 
    +J_g\frac{\rho_g}{\rho_0}\frac{1}{3}\frac{\partial \psi}{\partial \mathbf{H}}:\mathbb{\Omega}_M^{dev}:\overset{\circ}{\mathbf{L}}_M + J_g\frac{\rho_g}{\rho_0}\mathbf{F}_g^T\,(\nabla_0\times\mathbf{\Theta}):  \overset{\circ}{\mathbf{L}}_g
\end{align}
\end{footnotesize}

By denoting the remodelling configurational stress as $\boldsymbol{\zeta}=(1/3)\partial\psi/\partial \mathbf{H}$, an extended definition of the chemical potential $\mu$ conjugated to density variation can be given as:

\begin{align}\label{chempot}
    \mu &=\frac{\rho_g}{\rho_0}\frac{\partial \psi}{\partial (\rho_g/\rho_0)}+ \left[\mathbf{M}^{T}\boldsymbol{\zeta}\,\mathbf{M}^{-T}+\boldsymbol{\Sigma}:\mathbb{\Lambda}_M\right]:\mathbf{M}^{-1}\mathbf{\Omega}_\rho^{dev}\mathbf{M} \nonumber \\
    &\simeq \frac{\rho_g}{\rho_0}\frac{\partial \psi}{\partial (\rho_g/\rho_0)}+\left[\boldsymbol{\zeta}+\mathbf{M}^{-T}\boldsymbol{\Sigma}\mathbf{M}^{T}\right]:\mathbf{\Omega}_\rho^{dev} + h.o.t.
\end{align}

\begin{comment}
\begin{equation}\label{chempot}
    \mu=\frac{\rho_g}{\rho_0}\frac{\partial \psi}{\partial (\rho_g/\rho_0)}+ \left[\mathbf{M}^{T}\boldsymbol{\zeta}\,\mathbf{M}^{-T}+\boldsymbol{\Sigma}\right]:\mathbf{M}^{-1}\mathbf{\Omega}_\rho^{dev}\mathbf{M} -\mathcal{E}
\end{equation}
\end{comment}

The second term of this assumption involves the variation of free-energy density with respect to the rate of the anisotropic density contribution $\boldsymbol{\Upsilon}_g$. It is worth to highlight that, in \eqref{diss4}, the aliquot proportional to the trace of $\boldsymbol{\Sigma}$ is excluded from the hydrostatic chemical pressure $\mu$ since it constitutes a mutual work term in the light of the decomposition assumed for the effective chemo-mechanical stress $\boldsymbol{\mathcal{X}}$. Observing that the multiplicative decomposition of the total deformation gradient into its elastic and growth parts implies

\begin{equation}
    \dot{\mathbf{C}}=\dot{\overline{\mathbf{F}^T\mathbf{F}}} =  \overset{\circ}{\mathbf{L}}^T_g\mathbf{C}+\mathbf{C}\overset{\circ}{\mathbf{L}}_g+ \mathbf{F}_g^T\dot{\mathbf{C}}_e\mathbf{F}_g,
\end{equation}

one can directly identify from \eqref{diss4} the constitutive expression for the nominal stress:

\begin{equation}\label{stressass}
    \mathbf{S}= 2 J_g\frac{\rho_g}{\rho_0} \mathbf{F}_g^{-1}\frac{\partial \psi}{\partial \mathbf{C}_e}\mathbf{F}_g^{-T}.
\end{equation}

By accounting for this result, the dissipation \eqref{diss4} can be rearranged as  
\begin{footnotesize}
\begin{align}\label{diss5}
    & J_g\frac{\rho_g}{\rho_0}\boldsymbol{\mathcal{M}}_g\,:\overset{\circ}{\mathbf{L}}_g
    + J_g\frac{\rho_g}{\rho_0}\boldsymbol{\Sigma}:\overset{\circ}{\mathbf{L}}_g -J_g\frac{\rho_g}{\rho_0}\boldsymbol{\zeta}:\mathbb{\Omega}_M^{dev}:\overset{\circ}{\mathbf{L}}_g - J_g\frac{\rho_g}{\rho_0}\mathbf{F}_g^T\,(\nabla_0\times\mathbf{\Theta}):  \overset{\circ}{\mathbf{L}}_g  
    + J_g\,\frac{\rho_g}{\rho_0}\,\mu\,\frac{\dot{J}_g}{J_g}+ J_g\frac{\rho_g}{\rho_0}\Sigma_h\,\frac{\dot{\rho}_g}{{\rho}_g}\geq 0
\end{align}
\end{footnotesize}
%- \frac{\partial \psi}{\partial \mathbf{H}}:\mathbb{\Omega}_M^{dev}
%\boldsymbol{\Sigma}:\left[\frac{1}{3}\mathbf{I}+\frac{\rho_g}{\rho_0}\mathbf{\Omega}_\rho^{dev}\right]
where $\Sigma_h=(1/3)\,\boldsymbol{\Sigma}:\mathbf{I}$ the relation 

\begin{equation}\label{Mandel}
    \mathbf{C}\mathbf{S}=J_g\frac{\rho_g}{\rho_0}\, \left[\mathbf{F}_g^T\left(2 \mathbf{C}_e \frac{\partial \psi}{\partial \mathbf{C}_e} \right)\mathbf{F}_g^{-T}\right] = J_g\frac{\rho_g}{\rho_0}\, \left[\mathbf{F}_g^T\, \boldsymbol{\mathcal{M}}_e \,\mathbf{F}_g^{-T}\right] = J_g\frac{\rho_g}{\rho_0}\,\boldsymbol{\mathcal{M}}_g 
\end{equation}
has been introduced, by highlighting the reference Mandel type stress as configurational driving force involved in the inelastic growth process. This implies: 

%\frac{\partial \psi}{\partial \mathbf{H}}:\mathbb{\Omega}_M^{dev}

\begin{equation}\label{Sigma}
    \boldsymbol{\Sigma}=\boldsymbol{\zeta}:\mathbb{\Omega}_M^{dev}-\boldsymbol{\mathcal{M}}_g + \mathbf{F}_g^T\,(\nabla_0\times\mathbf{\Theta}),
\end{equation}

%in which, dually to the definition of chemical potential, the variation of free-energy density due to internal structural remodelling is included. 
Herein, the Mandel stress usually denoted as work-conjugate of the growth rate in many mechanical approaches based on constitutive growth laws not related to the mass balance \cite{ganghoffer2010eshelby, lamm2022macroscopic} is accompanied by the first term providing remodeling associated driving forces as well as nonlocal terms --proportional to a submacroscopic square length-- due to possible incompatibility in the grown configuration, this term vanishing in case of compatible and homogeneous growth. Also, the mass balance \eqref{genrate1} combined to relation \eqref{Sigma} implies that (part of) volumetric growth can be seen as resulting from the configurational stress momentum, the present results enriching for instance the constitutive assumptions introduced by Ganghoffer et al. in a purely mechanical context \cite{Goda_2016}. By virtue of \eqref{Sigma} and \eqref{Mandel} and by neglecting h.o.t., the chemical potential \eqref{chempot} can be further approximated as:

\begin{align}\label{chempot2}
    \mu \simeq \frac{\rho_g}{\rho_0}\frac{\partial \psi}{\partial (\rho_g/\rho_0)}+\left[\boldsymbol{\zeta}-\boldsymbol{\mathcal{M}}_e\right]:\mathbf{\Omega}_\rho^{dev} 
\end{align}

this expression highlighting the role of the effective configurational stress in directing material flow \cite{wu2001role}. In this regard, the effective driving stress $\boldsymbol{\mathcal{X}}$ can be rewritten as:

\begin{align}\label{Chi}
    \boldsymbol{\mathcal{X}} &=\mu\,\mathbf{I}+\boldsymbol{\Sigma}:\mathbb{\Lambda}_M \nonumber\\
    &\simeq \frac{\rho_g}{\rho_0}\frac{\partial \psi}{\partial (\rho_g/\rho_0)}\,\mathbf{I}+\{\left[\boldsymbol{\zeta}-\boldsymbol{\mathcal{M}}_e\right]:\mathbf{\Omega}_\rho^{dev}\}\,\mathbf{I}+ \boldsymbol{\zeta}:\mathbb{\Omega}_M^{dev}:\mathbb{\Lambda}_M-\boldsymbol{\mathcal{M}}_g:\mathbb{\Lambda}_M + \frac{\rho_0}{\rho_g J_g}\mathbf{F}_g^T\,(\nabla_0\times\mathbf{\Theta}):\mathbb{\Lambda}_M \nonumber\\
    & \simeq \left[\frac{\rho_g}{\rho_0}\frac{\partial \psi}{\partial (\rho_g/\rho_0)}\right]\,\mathbf{I} + \left[\boldsymbol{\zeta}-\boldsymbol{\mathcal{M}}_e\right]:\left[\mathbf{\Omega}_\rho^{dev}\otimes\mathbf{I}+\mathbb{\Omega}_M^{dev}\right]-\boldsymbol{\mathcal{M}}_g+ \mathbf{F}_g^T\,(\nabla_0\times\mathbf{\Theta}) 
\end{align}

After considering \eqref{Sigma}, dissipation rewrites as

\begin{align}\label{diss6}
    & J_g\frac{\rho_g}{\rho_0}\left[\mu\,\frac{\dot{J}_g}{J_g}+ \Sigma_h \frac{\dot{\rho}_g}{{\rho}_g}\right]\geq 0 , \quad \text{or} \nonumber\\
    & \mathcal{D}_{int}=\mu\,\rho_g\,\dot{J}_g+  \Sigma_h \,{J}_g\,{\dot{\rho}_g}\geq 0
\end{align}

that can be exploited to support the growth law \eqref{sys2}.

\begin{comment}
\begin{align}\label{diss6}
    & J_g\frac{\rho_g}{\rho_0}\left[\left(\frac{\rho_g}{\rho_0}\right)\mu\,\frac{\dot{J}_g}{J_g}+ \boldsymbol{\Sigma}:\left[\frac{1}{3}\mathbf{I}+\frac{\rho_g}{\rho_0}\mathbf{\Omega}_\rho^{dev}\right]\frac{\dot{\rho}_g}{{\rho}_g}\right]\geq 0 , \quad \text{or} \nonumber\\
    & \mathcal{D}_{int}=(\rho_0^{-1}\mu) \rho_g^2\,\dot{J}_g+ {J}_g \mathcal{F}_\rho \,{\dot{\rho}_g}\geq 0
\end{align}
\end{comment}

%where $\mathcal{F}_\rho=\boldsymbol{\Sigma}:\left[(1/3)\mathbf{I}+(\rho_g/\rho_0)\mathbf{\Omega}_\rho^{dev}\right]$.

\subsection{Maximum dissipation criterion}
A possible strategy to close the problem defined by the system \eqref{0prob2} is to assume that the mutual power density carried out by growth and densification \eqref{diss6} evolves by tending to maximize dissipation. Motivated by microstructural observations, the density of the system is assumed to evolve under free-stress conditions below certain carrying capacities that depend on interstitial spaces, diffusive potential and spontaneous reactions. Then, there exits a saturation density --say $\rho_g^{sat}$ (obtained as homogenized microstructural parameter at the macroscopic scale) below which the biological constituents (cells and other tissue components) can grow by filling the available interstitial spaces without involving volume addition or elastic interactions due to accumulation. In this latter case, densification would be the instead the result of a stressed environment. This implies that, once $\rho_g^{sat}$ is attained, density can evolve only by accounting for elastic effects, the natural growth of the body instead requiring the generation of new volume. From the dissipation \eqref{diss6}, let us consider the augmented dissipation \cite{Zuo_1998}:

%\mathcal{D}_{int}=\mu\,\rho_g\,\dot{J}_g+  \Sigma_h \,{J}_g\,{\dot{\rho}_g}

\begin{equation}
    \mathcal{D}_{int}^{a}=\mathcal{D}_{int}-\dot{\varpi}{J}_g(\rho_g-\rho_g^{sat})=\mu\,\rho_g\,\dot{J}_g+  \Sigma_h \,{J}_g\,{\dot{\rho}_g}-\dot{\varpi}{J}_g(\rho_g-\rho_g^{sat})
\end{equation}
where $\dot{\varpi}$ is an internal Lagrange multiplier translating the (workless) chemical population pressure generating after saturation event. It is thus subjected to the consistency conditions

\begin{equation}\label{persistence}
    \dot{\varpi}\,{\dot{\rho}_g}=0, \quad 
    \begin{cases}
     &\rho_g=\rho_g^{sat} \quad \text{and} \quad {\dot{\rho}_g}=0, \quad \dot{\varpi} \neq 0 \\
     &\rho_g<\rho_g^{sat} \quad, \quad \qquad \quad \quad \quad  \dot{\varpi} = 0
    \end{cases}
\end{equation}

Hill's maximum dissipation criterion with respect to the density then provides 

\begin{equation}
    \begin{cases}
        \frac{\partial \mathcal{D}_{int}^{a}}{\partial \rho_g}-\frac{\partial}{\partial t} \left( \frac{\partial \mathcal{D}_{int}^{a}}{\partial \dot{\rho}_g}\right)=0\\
        \dot{\varpi}\,{\dot{\rho}_g}=0
    \end{cases}
\end{equation}

For each of the cases determined by the condition \eqref{persistence}, and by further invoking the mass balance \eqref{genrate1}, the evolutions of density and volumetric growth can be decoupled in the two following sub-systems:

\begin{equation}\label{sys1w}
\text{Case 1}) \,\,
    \begin{cases}
        \frac{\dot{J}_g}{J_g}= \frac{\dot{\Sigma}_h}{\mu - \Sigma_h}\\
        \frac{\dot{\rho}_g}{{\rho}_g}+\frac{\dot{J}_g}{{J}_g}=\mathcal{R}_g\\
        \dot{\varpi}=0
    \end{cases},\qquad 
    \begin{aligned}
    &\rho_g<\rho_g^{sat}\quad\text{and} \quad \forall\,\mathcal{R}_g \quad \\ 
    &\qquad\qquad\cup \\
    &\rho_g=\rho_g^{sat}\quad\text{and} \quad \mathcal{R}_g<0
    \end{aligned}
\end{equation}

and 

\begin{equation}\label{sys2w}
\text{Case 2})  \,\,
    \begin{cases}
        \rho_g=\rho_g^{sat}\\ 
        \dot{\rho}_g=0, \\
        \frac{\dot{J}_g}{{J}_g}=\mathcal{R}_g\\
        \dot{\varpi}=\mu^{sat}\,\frac{\dot{J}_g}{J_g}
    \end{cases}, \quad \qquad\rho_g=\rho_g^{sat} \quad \text{and} \quad \mathcal{R}_g \geq 0
\end{equation}

This criterion is able to provide an additional equation that formally closes the growth problem. It is worth highlighting that in the saturated case (Case 2) it is obvious to observe that the bulk rate solely determines volumetric growth by recovering classical approaches, the generation of new material mediating a lagragean chemical pressure. Dually, when microstructural saturation has not been yet achieved or durng resorption (Case 1), the system can instead evolve by showing both density growth and volumetric growth, this latter one kindled by the possible presence of nonzero configurational stresses. In such a case, the auxiliary constitutive equation \eqref{sys1w}$_1$ can be advantageously exploited as additional equation to close the growth problem.

\section{Summary of the model}
In the light of the proposed strategy, the coupled model equations involving mechanical equilibrium, mass balance, growth and morphogenesis can be summarized as follows

\begin{equation} \label{SysDef1}
    \text{Case 1}) 
    \begin{cases}
        \nabla_0\cdot\mathbf{P}=\textbf{0}\\
        \frac{\dot{\rho}_g}{{\rho}_g}=\mathcal{R}_g-\frac{\dot{J}_g}{{J}_g}\\
        \frac{\dot{J}_g}{J_g}= \frac{\dot{\Sigma}_h}{\mu - \Sigma_h}\\
        \overset{\circ}{\mathbf{L}}_M=\mathbb{\Lambda}_M:\left[\boldsymbol{\mathcal{R}}_g^{dev} +\mathbf{M}^{-1}\,\mathbf{\Omega}_\rho^{dev}\,\mathbf{M}\,\frac{\dot{\rho}_g}{{\rho}_g}\right]\\
        \dot{\varpi}=0
        \end{cases} 
\end{equation}

and

\begin{equation}\label{SysDef2}
    \text{Case 2}) 
    \begin{cases}
        \nabla_0\cdot\mathbf{P}=\textbf{0}\\
        \dot{\rho}_g=0\\
        \frac{\dot{J}_g}{{J}_g}=\mathcal{R}_g\\
        \overset{\circ}{\mathbf{L}}_M=\mathbb{\Lambda}_M:\boldsymbol{\mathcal{R}}_g^{dev}\\
        \dot{\varpi}=\mu^{sat}\,\frac{\dot{J}_g}{J_g}
    \end{cases}
\end{equation}

which both give a closed system of 14 equations in 14 unknowns, considering the constitutive relations:

\begin{equation}\label{AllConst}
    \begin{cases}
        \mathbf{P}=J_g\frac{\rho_g}{\rho_0} \frac{\partial \psi}{\partial \mathbf{F}_e}\mathbf{F}_g^{-T}\\[1.25ex]
        \boldsymbol{\mathcal{M}}_g=\mathbf{F}_g^T\,\boldsymbol{\mathcal{M}}_e\,\mathbf{F}_g^{-T}=\mathbf{F}_g^T\left(2 \mathbf{C}_e \frac{\partial \psi}{\partial \mathbf{C}_e} \right)\mathbf{F}_g^{-T}\\[1.25ex]
        \boldsymbol{\zeta}=\frac{1}{3}\frac{\partial \psi}{\partial \mathbf{H}}\\[1.25ex]
        \boldsymbol{\Theta}=\frac{\partial\psi}{\partial(\nabla_0\times\mathbf{F}_g)}\\[1.25ex]
        \mu \simeq \frac{\rho_g}{\rho_0}\frac{\partial \psi}{\partial (\rho_g/\rho_0)}+\left[\boldsymbol{\zeta}-\boldsymbol{\mathcal{M}}_e\right]:\mathbf{\Omega}_\rho^{dev} \\[1.25ex]
        \boldsymbol{\Sigma}=\left[\boldsymbol{\zeta}:\mathbb{\Omega}_M^{dev}-\boldsymbol{\mathcal{M}}_g+ \mathbf{F}_g^T\,(\nabla_0\times\mathbf{\Theta})\right]\\[1.25ex]
        \boldsymbol{\mathcal{X}}\simeq \left[\frac{\rho_g}{\rho_0}\frac{\partial \psi}{\partial (\rho_g/\rho_0)}\right]\,\mathbf{I} + \left[\boldsymbol{\zeta}-\boldsymbol{\mathcal{M}}_e\right]:\left[\mathbf{\Omega}_\rho^{dev}\otimes\mathbf{I}+\mathbb{\Omega}_M^{dev}\right]-\boldsymbol{\mathcal{M}}_g+ \mathbf{F}_g^T\,(\nabla_0\times\mathbf{\Theta})\\[1.25ex]
        \mathbf{Q}=- \boldsymbol{\kappa}\cdot\left(\frac{\boldsymbol{\mathcal{X}}}{\rho_0}\right)\cdot \nabla_0\\
    \end{cases}
\end{equation}

\section{Conclusions}
%%potremmo usare queste conclusioni ridotte per ArXiv
%%In these notes, we have developed a novel theoretical framework for describing anisotropic growth, establishing an inherent connection between the evolution of mass and geometry. Through the generalization of mass balance and the identification of suitable equations for tracing back the spontaneous dynamics of tissue growth and morphogenesis, the proposed approach opens avenues for applications in growth, remodelling and morphogenesis, by avoiding the need of specifying the structure of the growth tensor or \textit{ad hoc} constitutive hypotheses for the growth law and by using measurable and physically meaningful parameters. While challenges such as computational complexity remain, the developed model could be employed to explore more sophisticated growth scenarios, its generality potentially contributing to further bridging the gap between biomechanics and mechanobiology perspectives towards a unified approach by melding the essential mechanical and biological features.

%notes
Morphoelasticity represents one of the most consolidated theories to the modern descriptions of growth, remodelling and morphogenesis phenomena. However, many challenging points remain still partly unexplored, mainly involving \textit{i}) the identification of suitable kinematic growth laws respecting a general principle common to all the living systems, thus avoiding phenomenological and \textit{a priori} particularized tensor structures, and \textit{ii}) how the local growth and morphogenesis couple to shape, highlighting the explicit role of mechanics and diffusion in enhancing morphogenic gradients and chemical pathways that lead to the shaping of growing systems \cite{goriely2017mathematics, ambrosi2019growth, lewicka2022geometry}. To the latter ones, the possibility of defining a local descriptor of the morphogenesis without resorting to stress-mediated events occurring at a more global scale (as in the case growth-induced instability) seem poorly investigated. 
To contribute at least in part respond to these open issues, the presented approach aims to unify the leading aspects involved in mechanobiology processes, by connecting accretion, shape change and micro-structural development through the sole mass conservation equation. Under the hypothesis that mass balance should represent the essential dynamic constraint of the growth problem and should thus perform as the sole exhaustive regulator of the biological, geometrical and constitutive transformations possibly undergone by the living material, we introduced a generalization of the scalar mass conservation in order to derive a consistent growth-law that avoids the need of either specifying the structure of the growth tensor or introducing constitutive relations and prevents the uncoupling of geometry and mass. 
As well-known, mass balance already provides complete phenomenological information about the leading species' proliferation, decay and mutual interactions --in the form of competition/cooperation dynamics and plasticity (mutation) terms-- as well as about mass flow and re-organization through \textit{ad hoc} diffusion mechanisms \cite{fraldi2018cells, bernard2024modelling}. All these phenomena that can provide feedback terms with the chemical and mechanical environment within a full-coupled and multi-physic view, by providing experimentally informed inhomogeneous rates that depend on the mechanical stress and/or massless chemicals' concentrations. 
Through the presented framework of anisotropic growth, intrinsic living rates are combined to all the relevant mechanical and transport properties, so contributing to the overall material evolution. The hypothesis that mass and geometry evolve synergistically  let to show how flow and source terms, besides contributing to bulk changes, can be actually used to potentially trace back the process of shape formation though introducing a local descriptor of the morphogenesis capable of driving shape changes also in absence of configurational switches.  Thanks to its generality, the proposed theoretical strategy can help address some of the above challenges, offering avenues to explore complex scenarios of growth, remodeling, and morphogenesis. This potentially contributes to bridging the gap between biomechanics and mechanobiology, advancing toward a unified perspective.

\section*{Acknowledgements}

ARC, SP and AC acknowledge the financial support under the National Recovery and Resilience Plan (NRRP), Mission 4, Component 2, Investment 1.1, Call for tender No. 1409 published on 14.9.2022 by the Italian Ministry of University and Research (MUR), funded by the European Union – NextGenerationEU– Projects: \textit{i})  P2022M3KKC MECHAVERSE – CUP E53D23017310001 (A.R.C.), \textit{ii}) P2022AC8H4 MEDUSA – CUP E53D23017050001 (S.P.), and \textit{iii}) P2022MXCJ2 (A.C.), Grant Assignment Decrees No. 1385 and No. 1379 adopted on 1.9.2023 by the Italian Ministry of University and Research (MUR)). A.C. has been supported by the Project of National Relevance PRIN2022 grant no 2022XLBLRX funded by the Italian MUR. 
MF acknowledges the financial support under the National Recovery and Resilience Plan (NRRP), Mission 4, Component 2, Investment 1.1, Call for tender No. 104 published on 2.2.2022 by the Italian Ministry of University and Research (MUR), funded by the European Union – NextGenerationEU– Project Title 2022ATZCJN AMPHYBIA – CUP E53D23003040006 - Grant Assignment Decree No. 961 adopted on 30.06.2023 by the Italian Ministry of University and Research (MUR).

\newpage
\section*{Appendix}\label{AppA}
\renewcommand{\theequation}{A.\arabic{equation}}
\subsection*{Considerations on fabric tensors}
%Paragrafo 1 - Definizione
\paragraph{Relation between material inhomogeneity and fabric remodelling.} Fabric tensors are commonly used to quantify the degree of anisotropy of material microstructure, obtained by refining the description of a certain representative volume element at a sub-macroscopic scale. They are operatively computed by means of different techniques, for instance based on the Mean-Intercept-Length (MIL) approaches or the Gradient Mean Intercept Length (MIL) or Gradient Structural Tensors (GST) \cite{moreno2014techniques}.
In these morphology-based approaches, when the inhomogeneity of the material is taken into explicit account, 
fabric tensors are computed by inspecting the material architecture by taking into account for density inhomogeneity along selected directions, thus by considering the value of the linear density and of its spatial variability. Considering a certain geometric direction $\mathbf{n}$, the material is inspected by investigating an operator of the type

\begin{equation}
    \varrho(\mathbf{n})\mathbf{n}\otimes \varrho(\mathbf{n})\mathbf{n}=\varrho^2(\mathbf{n})\mathbf{N}
\end{equation}

where $\varrho$ is the linear density along the direction $\mathbf{n}$, defined such that $\varrho(\mathbf{n})=\varrho(-\mathbf{n})$. At a certain macroscopic point $\mathbf{X}$ of the body, a first-order approximation of the density over a sub-macroscopic representative volume leads to write

\begin{equation}
    \varrho(\mathbf{n},\mathbf{X})\simeq \tilde{\varrho}(\mathbf{X})+ \ell\,\nabla\tilde{\varrho}(\mathbf{X})\cdot\mathbf{n}=
\end{equation}
where $\ell$ is the spatial resolution of material inspection and $\tilde{\varrho}$ is the density in the central material point. Therefore, the square of the linear density reads as
\begin{align}
    \varrho(\mathbf{n},\mathbf{X})^2 &=\tilde{\varrho}^2\left[1+ 2\ell \frac{\nabla\tilde{\varrho}}{\tilde{\varrho}}\cdot\mathbf{n}+\ell^2\left(\frac{\nabla\tilde{\varrho}}{\tilde{\varrho}}\cdot\mathbf{n}\right)\left(\frac{\nabla\tilde{\varrho}}{\tilde{\varrho}}\cdot\mathbf{n}\right) \right]=\nonumber\\
    %%%
    &=\tilde{\varrho}^2\mathbf{n}\cdot\left[\mathbf{I}+\ell\left(\mathbf{n}\otimes\frac{\nabla\tilde{\varrho}}{\tilde{\varrho}}+\frac{\nabla\tilde{\varrho}}{\tilde{\varrho}}\otimes\mathbf{n}\right)+ \ell^2\left(\frac{\nabla\tilde{\varrho}}{\tilde{\varrho}}\otimes\frac{\nabla\tilde{\varrho}}{\tilde{\varrho}}\right)\right]\cdot\mathbf{n}
\end{align}

Odd terms are avoided by considering 

\begin{align}\label{nDn}
    {\varrho}_e^2=\frac{\varrho(\mathbf{n},\mathbf{X})^2+\varrho(-\mathbf{n},\mathbf{X})^2}{2}=\tilde{\varrho}^2\,\mathbf{n}\cdot\left[\mathbf{I}+\ell^2\left(\frac{\nabla\tilde{\varrho}}{\tilde{\varrho}}\otimes\frac{\nabla\tilde{\varrho}}{\tilde{\varrho}}\right)\right]\cdot\mathbf{n}= \tilde{\varrho}^2\, \mathbf{D}:\mathbf{N},
\end{align}

where the tensor $\mathbf{D}$ in square brackets measures the material variability of the quadratic stereological density within a neighborhood of size $\ell$. In the natural state, considering that the linear density quantity ${\varrho}_e$ scales as $\rho_g^{1/3}$, one obtains an expression for the the sterelogical density distribution as 

\begin{align}\label{sterdens}
    {\varrho}_e&=\rho_g^{1/3}\,\left[\mathbf{I}+\frac{\ell^2}{9}\left(\frac{\nabla_g\rho_g}{\rho_g}\otimes\frac{\nabla_g\rho_g}{\rho_g}\right)\right]^\frac{1}{2}:\mathbf{N}= \rho_g^{1/3}\,\left[\mathbf{I}+2\ell^2_g\left(\frac{\nabla_g\rho_g}{\rho_g}\otimes\frac{\nabla_g\rho_g}{\rho_g}\right)\right]^\frac{1}{2}:\mathbf{N},
    \Rightarrow   \nonumber\\
    {\varrho}_e&\mathbf{N}=\rho_g^{1/3}\mathbf{D}^\frac{1}{2}, \nonumber\\
    \mathbf{D}&=\mathbf{I}+2\ell^2_g\boldsymbol{\xi}_g, \qquad \boldsymbol{\xi}_g=\rho_g^{-2}{\nabla_g\rho_g}\otimes{\nabla_g\rho_g}
\end{align}

A comparison then gives a direct estimation of the anisotropy density $\boldsymbol{\Upsilon}_g$ introduced in \eqref{dmg2}, by posing

\begin{align}\label{Upsilon}
   \boldsymbol{\Upsilon}_g=\frac{1}{D^{1/6}}\mathbf{D}^\frac{1}{2}, \quad D=\det \mathbf{D} 
\end{align}

in which $\ell_g$ is a sub-macroscopic length in the grown configuration that weights the inhomogeneity of the material. On the other hand, the density distribution tensor in equation \eqref{sterdens} helps to also define many morphology-based methods for the computation of fabric tensors, including the Mean Intercept Length (MIL) or Gradient Structural Tensors (GST) \cite{moreno2014techniques}. Based on these approaches, Cowin \cite{cowin1992evolutionary, cowin1989identification, jemiolo1997fabric} defined a fabric tensor proportional to the inverse of the square root of the MIL tensor, i.e. $\mathbf{H}\propto \mathbf{D}^{-1/2}$. Therefore, by starting from equation \eqref{sterdens}, a first order approximation and opportune normalization allow us to define the following expression for the fabric tensor

\begin{align}\label{fabricHD}
    &\mathbf{H}=\frac{1}{3}\left[\mathbf{I}-\ell^2_g\,\mathbb{P}^{dev}:\left(\frac{\nabla_g\rho_g}{\rho_g}\otimes\frac{\nabla_g\rho_g}{\rho_g}\right)\right]
\end{align}

where $\mathbb{P}^{dev}=[\mathbb{I}-(1/3)(\mathbf{I}\otimes\mathbf{I})]$ is the deviatoric projector. For the present evolutionary problems, it is convenient to express this tensor with respect to reference coordinates as

\begin{align}\label{Hrho}
    &\mathbf{H}=\tilde{\mathbf{H}}(\mathbf{M},\rho_g,\nabla_0\rho_g\otimes\nabla_0\rho_g)=\frac{1}{3}\left[\mathbf{I}-\ell^2_0\,tr(\mathbf{C}_M)\,\mathbb{P}^{dev}:\,\mathbf{M}^{-{T}}\,\boldsymbol{\xi}\,\mathbf{M}^{-1}\right],
\end{align}
where the tensor $\boldsymbol{\xi}$ denotes the tensor $\boldsymbol{\xi}_g$ in the reference coordinates:

\begin{align}\label{xieq}
     &\boldsymbol{\xi}=\frac{\nabla_0\rho_g}{\rho_g}\otimes\frac{\nabla_0\rho_g}{\rho_g}=\rho_g^{-2}\boldsymbol{\nabla}_0\rho_g^2
\end{align}

\paragraph{Relation between the rates of density anisotropy tensor and fabric tensor}. The definitions of tensors $\boldsymbol{\Upsilon}_g$ and $\mathbf{H}$ introduced respectively in equations \eqref{sterdens} and \eqref{fabricHD} motivate the direct connection between the rates $\boldsymbol{\Upsilon}_g^{-1}\dot{\boldsymbol{\Upsilon}_g}$ and $\dot{\mathbf{H}}$ assumed in the text. To demonstrate this, by neglecting for now the possible contributions of growth spins for both tensor, let us start from performing the rates of square tensor $\boldsymbol{\Upsilon}_g^2=D^{-1/3}\mathbf{D}$ as

\begin{align}\label{DUps}
    &2\boldsymbol{\Upsilon}_g\dot{\boldsymbol{\Upsilon}_g}=D^{-1/3}\dot{\mathbf{D}}-\frac{1}{3}D^{-4/3}(\mathbf{D}^{-1}:\dot{\mathbf{D}})\mathbf{D}=\boldsymbol{\Upsilon}_g^2\left[\mathbf{D}^{-1}\dot{\mathbf{D}}-\frac{1}{3}(\mathbf{D}^{-1}:\dot{\mathbf{D}})\mathbf{I}\right]+ h.o.t., \Rightarrow\nonumber\\
    &\boldsymbol{\Upsilon}_g^{-1}\dot{\boldsymbol{\Upsilon}_g}=\frac{1}{2}\left[\mathbf{D}^{-1}\dot{\mathbf{D}}-\frac{1}{3}(\mathbf{D}^{-1}:\dot{\mathbf{D}})\mathbf{I}\right] + h.o.t.=\frac{1}{2}\mathbb{P}^{dev}:(\mathbf{D}^{-1}\dot{\mathbf{D}})+h.o.t.
\end{align}

On the other hand, under the same hypotheses, direct derivation of relation \eqref{fabricHD} gives

\begin{align}\label{DHg}
    &\dot{\mathbf{H}}=-\frac{1}{3}\,\mathbb{P}^{dev}:\dot{\overline{\ell^2_g\,\boldsymbol{\xi}_g}}= -\frac{1}{6}\,\mathbb{P}^{dev}:\dot{\mathbf{D}} + h.o.t. 
\end{align}

Expressions \eqref{DUps} and \eqref{DHg} can be compared directly with the help of the Sherman-Morrison formula (\textbf{CIT}) applied to the tensor $\mathbf{D}$ in equation \eqref{sterdens}:

\begin{align}
    [\mathbf{A}+\mathbf{u}\otimes\mathbf{v}]^{-1}&=\mathbf{A}^{-1}-\frac{\mathbf{A}^{-1}\cdot(\mathbf{u}\otimes\mathbf{v})\cdot\mathbf{A}^{-1}}{1+\mathbf{v}^T\cdot\mathbf{A}^{-1}\cdot\mathbf{u}} \quad \Rightarrow \nonumber\\
    \mathbf{D}^{-1}&=\mathbf{I}-\frac{2\ell^2_g\boldsymbol{\xi}_g}{1+2\ell^2_g\,tr(\boldsymbol{\xi}_g)}, \qquad \boldsymbol{\xi}_g=\rho_g^{-2}{\nabla\rho_g}\otimes{\nabla\rho_g}
\end{align}

By virtue of this relation and considering small spatial resolutions, relations \eqref{DUps} and \eqref{DHg} can be approximated as follows

\begin{align}
    &\boldsymbol{\Upsilon}_g^{-1}\dot{\boldsymbol{\Upsilon}_g}\simeq\frac{1}{2}\left[\dot{\mathbf{D}}-\frac{1}{3}(\mathbf{I}:\dot{\mathbf{D}})\mathbf{I}\right] + h.o.t., \quad \text{and}\nonumber\\
    &\dot{\mathbf{H}}\simeq -\frac{1}{6}\left[\dot{\mathbf{D}}-\frac{1}{3}(\mathbf{I}:\dot{\mathbf{D}})\mathbf{I}\right] + h.o.t.
\end{align}

from which follows the ansatz $\boldsymbol{\Upsilon}_g^{-1}\dot{\boldsymbol{\Upsilon}_g}\simeq -3 \dot{\mathbf{H}}$ introduced in equation \eqref{trHY2} for the charaterization of the generalized mass rate. 
In addition, a comparison between the expressions of $\boldsymbol{\Upsilon}_g$ and $\mathbf{H}$ in equations \eqref{sterdens} and \eqref{fabricHD} let to find the following relation: 
\begin{equation}
    \log \boldsymbol{\Upsilon}_g = \frac{1}{2}\left(\mathbf{I}-3\mathbf{H}\right)
\end{equation}

\paragraph{Estimation of the remodelling rate coefficients.} From the expression of the reference fabric tensor \eqref{Hrho}, the remodelling rate coefficients of equation \eqref{Hdot} can be evaluated by considering the variation of $\mathbf{H}$ within a control inspection volume centred in a macro-point $\mathbf{X}$:

\begin{align}
    \frac{1}{|\Omega_y|}\delta\int_\Omega\,\mathbf{H} d\omega = \frac{1}{|\Omega_y|}\int_{\Omega_y}\, &\left[\frac{\partial \mathbf{H}}{\partial \mathbf{M}}:(\mathbf{M}\overline{\otimes}\mathbf{I}):\mathbf{M}^{-1}\delta\mathbf{M}\right]_{(\mathbf{X},y)}+\\
    &+\left\{\frac{\rho_g}{\rho_0}\,\left[\rho_0\,\left(\frac{\partial \mathbf{H}}{\partial \rho_g}-\frac{\partial \mathbf{H}}{\boldsymbol{\nabla}\rho_g^2}:2\,\text{sym}(\nabla\rho_g\otimes \nabla)\right) \right]\,\frac{\delta{\rho}_g}{\rho_g}\right\}_{(\mathbf{X},y)} \, \, d^3\mathbf{y} + \notag\\
    &+\frac{1}{|\Omega_y|}\int_{\partial \Omega_y}2\left(\mathbf{n}\cdot\frac{\partial \mathbf{H}}{\boldsymbol{\nabla}\rho_g^2}\cdot\nabla\rho_g\, \delta\rho_g\right)_{(\mathbf{X},y)}\, \, d^2\mathbf{y} 
\end{align}

Application of the mean value theorem with respect to the central macro-point $\mathbf{X}$ and by neglecting a variation of the bulk density over the control closed surface of the $\mathbf{X}$-neighborhood gives:

\begin{align}\label{Hdotfull}
    \dot{\mathbf{H}} = &\left[\frac{\partial \mathbf{H}}{\partial \mathbf{M}}:(\mathbf{M}\overline{\otimes}\mathbf{I})\right]:\mathbf{M}^{-1}\dot{\mathbf{M}}+\frac{\rho_g}{\rho_0}\,\left[\rho_0\,\left(\frac{\partial \mathbf{H}}{\partial \rho_g}-\frac{\partial \mathbf{H}}{\boldsymbol{\nabla}_0\rho_g^2}:2\,\text{sym}(\nabla_0\rho_g\otimes \nabla_0)\right) \right]\,\frac{\dot{\rho}_g}{\rho_g}=\notag\\
    &=\frac{1}{3}\mathbb{\Omega}_M^{dev}:\overset{\circ}{\mathbf{L}}_M+\frac{1}{3}\boldsymbol{\Omega}_\rho^{dev}\,\frac{\dot{\rho}_g}{\rho_g}
\end{align}

Through the use of equations \eqref{Hdotfull} and \eqref{xieq}, application of the chain rule let to determine the remodelling rate coefficients of the explicit expression \eqref{Hrho}. More precisely, from evaluating

 \begin{footnotesize}
\begin{align}
\frac{\partial \mathbf{H}}{\partial \rho_g}=\frac{\partial \mathbf{H}}{\partial \boldsymbol{\xi}}:\frac{\partial \boldsymbol{\xi}}{\partial \rho_g}&=-\frac{\ell_0^2\, tr(\mathbf{C}_M)}{3}\,\mathbb{P}^{dev}:\left(\mathbf{M}^{-T}\overline{\underline{\otimes}}\,\mathbf{M}^{-T}\right):\left(\frac{-2}{\rho_g}\right)\boldsymbol{\xi}
=\frac{2\ell_0^2\, tr(\mathbf{C}_M)}{3\rho_g}\,\mathbb{P}^{dev}:\mathbf{M}^{-T}\boldsymbol{\xi}\mathbf{M}^{-1}
\end{align}
\end{footnotesize}

and

\begin{align}
\frac{\partial \mathbf{H}}{\boldsymbol{\nabla}_0\rho_g^2}:(2\nabla_0\rho_g\otimes \nabla_0)_{s}&=-\frac{\ell_0^2\, tr(\mathbf{C}_M)}{3 \rho_g^2}\,\mathbb{P}^{dev}:\left(\mathbf{M}^{-T}\overline{\underline{\otimes}}\,\mathbf{M}^{-T}\right):\mathbb{I}:(2\nabla_0\rho_g\otimes \nabla_0)_{s}=\notag\\
&=-\frac{2\ell_0^2\, tr(\mathbf{C}_M)}{3 \rho_g^2}\,\mathbb{P}^{dev}:[\mathbf{M}^{-T}\,(\nabla_0\rho_g\otimes \nabla_0)_{s}\,\mathbf{M}^{-1}]
\end{align}

Therefore, one obtains:
\begin{equation}\label{OmegaRhodef}
    \boldsymbol{\Omega}_\rho^{dev}=2\ell_0^2\, tr(\mathbf{C}_M) \mathbb{P}^{dev}:\left[\mathbf{M}^{-T}\left(\boldsymbol{\xi}+\frac{1}{\rho_g}(\boldsymbol{\nabla}_0\rho_g)_{s}\right)\mathbf{M}^{-1}\right]
\end{equation}

This latter coefficient being directly involved in considering the remodelling of the material due density rearrangement and in the morphogenesis evolution law (see equations  \eqref{LMeq} and \eqref{sys2}$_3$). The overall fabric evolution \eqref{Hdotfull} also depends on geometric changes. In a similar way, the fourth rank rate coefficient related to the geometric deformation can be derived by considering:

 \begin{footnotesize}
\begin{align}
    \frac{\partial \mathbf{H}}{\partial \mathbf{M}}&=-\frac{\ell_0^2}{3}\left[(\mathbb{P}^{dev}:\mathbf{M}^{-T}\boldsymbol{\xi}\mathbf{M}^{-1})\otimes 2
    \mathbf{M} + tr(\mathbf{C}_M)\, \mathbb{P}^{dev}:\left(\frac{\partial \mathbf{M}^{-T}}{\partial \mathbf{M}}\,\boldsymbol{\xi}\mathbf{M}^{-1}+\mathbf{M}^{-T}\,\boldsymbol{\xi}\,\frac{\partial \mathbf{M}^{-1}}{\partial \mathbf{M}}\right)\right]=\notag\\
    &=\frac{\ell_0^2}{3}\mathbb{P}^{dev}:\left[tr(\mathbf{C}_M)\,(\mathbf{M}^{-T}\boldsymbol{\xi}\mathbf{M}^{-1}\underline{\otimes}\,\mathbf{M}^{-T}+\mathbf{M}^{-T}\boldsymbol{\xi}\mathbf{M}^{-1}\overline{\otimes}\,\mathbf{M}^{-1})-2(\mathbf{M}^{-T}\boldsymbol{\xi}\mathbf{M}^{-1}\otimes \mathbf{M})\right]
\end{align}
\end{footnotesize}

 \begin{footnotesize}
\begin{align}\label{OmegaMdef}
    \mathbb{\Omega}_M^{dev} = \ell_0^2\,\mathbb{P}^{dev}:
    &\left\{tr(\mathbf{C}_M) \left[(\mathbf{M}^{-T}\boldsymbol{\xi}\mathbf{M}^{-1})\,\underline{\otimes}\,\mathbf{M}^{-T}\mathbf{M}+\mathbf{M}^{-T}\boldsymbol{\xi}\, \overline{\otimes}\, \mathbf{M}^{-1}\right] - 2\left[(\mathbf{M}^{-T}\boldsymbol{\xi}\mathbf{M}^{-1}\otimes \mathbf{C}_M) \right]\right\}
\end{align}
\end{footnotesize}

\subsection*{Generalized mass rate}
In Section \ref{sec:genmass} we have expressed the mass element as the determinant of generalized mass vectors along three $i$-th mutual directions, given by:

\begin{equation}\label{dmgapp}
    dm_g=\det [d\mathbf{m}_{g,i}], \quad d\mathbf{m}_{g,i}=\rho_g^{1/3}\,\boldsymbol{\Upsilon}_g\cdot d\mathbf{X}_{g,i}= \boldsymbol{y}_g\cdot\mathbf{F}_g\cdot d\mathbf{X}_{0,i}=\boldsymbol{\rho}_g\cdot d\mathbf{X}_{0,i}
\end{equation}

This means that we can establish for the growth and densification tensor $\boldsymbol{\rho}_g$ (with a little abuse of notation) as:

\begin{equation}
    \boldsymbol{\rho}_g=\boldsymbol{y}_g\cdot\mathbf{F}_g=d\mathbf{m}_{g,i}\otimes d\mathbf{X}^{0,i}
\end{equation}

The variation of mass in time writes as 

\begin{align}\label{bmapp}
    \frac{d}{dt}\int\, dm_g &= \frac{d}{dt} \int_{V^0} \det[\boldsymbol{\rho}_g] dV^0= \int_{V^0}\frac{d}{dt}\det[\boldsymbol{\rho}_g]\, dV^0 \nonumber\\
    &= \int_{V^0} \left(\boldsymbol{\rho}_g^{-T}:\dot{\boldsymbol{\rho}}_g\right) \det[\boldsymbol{\rho}_g]\, dV^0=\int tr\left(\boldsymbol{\rho}_g^{-1}\,\dot{\boldsymbol{\rho}}_g\right) dm_g = \int \mathcal{R}_g dm_g
\end{align}

To find an expression for the generalized mass rate tensor $\boldsymbol{\mathcal{R}}_g$ such that $tr\left(\boldsymbol{\rho}_g^{-1}\,\dot{\boldsymbol{\rho}}_g\right)=\mathcal{R}_g$, we start from the time variation of mass vector, which reads

\begin{equation}
    \dot{d\mathbf{m}}_{g,i}= \dot{\boldsymbol{\rho}}_g\cdot d\mathbf{X}_{0,i}, \quad \dot{\boldsymbol{\rho}}_g=\dot{d\mathbf{m}}_{g,i}\otimes d\mathbf{X}^{0,i}
\end{equation}

then

\begin{equation}\label{par1}
   \boldsymbol{\rho}_g^{-1}\,\dot{\boldsymbol{\rho}}_g  = (d\mathbf{m}^{g,i}\cdot\dot{d\mathbf{m}}_{g,i})\,d\mathbf{X}_{0,i}\otimes d\mathbf{X}^{0,i}, \quad d\mathbf{X}^{0,i}\cdot(\boldsymbol{\rho}_g^{-1}\,\dot{\boldsymbol{\rho}}_g)\cdot\,d\mathbf{X}_{0,i}=(d\mathbf{m}^{g,i}\cdot\dot{d\mathbf{m}}_{g,i})
\end{equation}

This expression shows that the material rate tensor $\boldsymbol{\rho}_g^{-1}\,\dot{\boldsymbol{\rho}}_g$ weights the rate of each material vector. Indeed, the scalar product $(d\mathbf{m}^{g,i}\cdot\dot{d\mathbf{m}}_{g,i})$ can be written as

\begin{align}\label{par2}
    d\mathbf{m}^{g,i}\cdot\dot{d\mathbf{m}}_{g,i} &= d\mathbf{X}^{0,i}\cdot\left[\mathbf{F}_g^{-1}\boldsymbol{y}_g^{-1}(\dot{\boldsymbol{y}}_g\mathbf{F}_g+\boldsymbol{y}_g\dot{\mathbf{F}}_g)\right]\cdot\,d\mathbf{X}_{0,i}=\nonumber\\
    &=d\mathbf{X}^{0,i}\cdot\mathbf{F}_g^{-1}\cdot\left[\boldsymbol{y}_g^{-1}\dot{\boldsymbol{y}}_g+\dot{\mathbf{F}}_g\mathbf{F}_g^{-1}\right]\cdot \mathbf{F}_g\cdot\,d\mathbf{X}_{0,i}= d\mathbf{X}^{0,i}\cdot\mathbf{F}_g^{-1}\cdot\boldsymbol{r}_g\cdot\mathbf{F}_g\cdot\,d\mathbf{X}_{0,i}
\end{align}

Herein, the tensor $\boldsymbol{r}_g$ represents the rate performed on the grown configuration, given by the sum of the natural geometrical rate and the densification rate, which is pulled-back from the grown-and-densified configuration to the grown configuration. Direct comparison of \eqref{par1} and \eqref{par2} gives

\begin{equation}
    \boldsymbol{\rho}_g^{-1}\,\dot{\boldsymbol{\rho}}_g =   \boldsymbol{\mathcal{R}}_g = \mathbf{F}_g^{-1}\cdot\boldsymbol{r}_g\cdot\mathbf{F}_g, \qquad  tr(\boldsymbol{\mathcal{R}}_g)=tr(\boldsymbol{r}_g)
\end{equation}

The balance of mass \eqref{bmapp} can be detailed as usual by accounting for fluxes and source/sink terms:

\begin{align}
    &\int \mathcal{R}_g dm_g = - \int_{\partial V^g} \mathbf{q}_g\cdot\hat{\boldsymbol{\nu}}_g\,dS^g + \int_{V^g} \Gamma \rho_g dV^g, \Rightarrow \nonumber \\
    &\int \boldsymbol{r}_g dm_g = - \int_{\partial V^g} \mathbf{q}_g \otimes \hat{\boldsymbol{\nu}}_g\,dS^g + \int_{V^g} \mathbf{H}\, \Gamma \rho_g dV^g =  - \int_{V^g} \mathbf{q}_g \otimes \nabla_g\,dV^g + \int_{V^g} \mathbf{H}\, \Gamma \rho_g dV^g \nonumber \\
    &\int_{V^0} \boldsymbol{r}_g \rho_g\,J_g\,dV^0 = - \int_{V^0} J_g\mathbf{q}_g \otimes \nabla_0\cdot\mathbf{F}_g^{-1}\,dV^0 + \int_{V^0} \mathbf{H}_g \Gamma \rho_g\,J_g\,dV^0
\end{align}

where the first term related to fluxes provides a direct generalization of the Nanson's formula in combination with the Gauss theorem, while $\mathbf{H}$ denotes the fabric tensor of the newly grown material element in the natural state (see equation \eqref{Hrho}). Hence

\begin{align}
    &\boldsymbol{r}_g = - \frac{1}{\rho_g\,J_g} (J_g\mathbf{q}_g \otimes \nabla_0\cdot\mathbf{F}_g^{-1}) + \mathbf{H}\, \Gamma, \Rightarrow \nonumber\\
    &\boldsymbol{\mathcal{R}}_g = \mathbf{F}_g^{-1}\cdot\boldsymbol{r}_g\cdot\mathbf{F}_g= - \frac{1}{\rho_g\,J_g} (\mathbf{Q}\otimes \nabla_0) + \mathbf{M}^{-1}\cdot\boldsymbol{H}\cdot\mathbf{M}\, \Gamma, 
\end{align}

which thus returns the growth rate form introduced in Section \ref{sec:genmass}, under the position $\mathbf{Q}=J_g\mathbf{q}_g\,\mathbf{F}_g^{-T}$.

\bibliographystyle{spphys}
\bibliography{MorphoBib}

\end{document}